\documentclass[traditabstract]{aa}

\usepackage{graphicx}
\usepackage{txfonts}
\usepackage{rotating}

\begin{document}

\title{Constraining the cometary flux through the asteroid belt during~the~late heavy bombardment}
\authorrunning{M.~Bro\v z et al.}

\subtitle{}

\author{M.~Bro\v z\inst{1}, A.~Morbidelli\inst{2}, W.F.~Bottke\inst{3}, J.~Rozehnal\inst{1}, D.~Vokrouhlick\'y\inst{1}, D.~Nesvorn\'y\inst{3}}

\institute{Institute of Astronomy, Charles University, Prague, V Hole\v sovi\v ck\'ach 2, 18000 Prague 8, Czech Republic,\\
e-mail:
mira@sirrah.troja.mff.cuni.cz,
rozehnal@observatory.cz,
davok@cesnet.cz
\and
Observatoire de la C\^ote d'Azur, BP 4229, 06304 Nice Cedex 4, France,
e-mail: morby@oca.eu
\and
Department of Space Studies, Southwest Research Institute, 1050 Walnut St., Suite 300, Boulder, CO 80302, USA,\\
e-mail: bottke@boulder.swri.edu,
davidn@boulder.swri.edu
}

\date{Received ???; accepted ???}

% \abstract{}{}{}{}{} 
% 5 {} token are mandatory
 
\abstract{%
In the Nice model, the late heavy bombardment (LHB) is related
to an orbital instability of giant planets which causes
a fast dynamical dispersion of a transneptunian cometary disk.
% (Gomes et al. 2005).
We study effects produced by these hypothetical cometary projectiles
on main-belt asteroids. In particular, we want to check whether the observed collisional
families provide a lower or an upper limit for the cometary flux during the LHB.

We present an updated list of observed asteroid families as identified
in the space of synthetic proper elements by the hierarchical clustering method,
colour data, albedo data and dynamical considerations and we estimate their physical parameters.
We selected $12$~families which may be related to the LHB according to their dynamical ages.
We then used collisional models and N-body orbital simulations to gain insight into
the long-term dynamical evolution of synthetic LHB families over $4\,{\rm Gyr}$.
We account for
the mutual collisions between comets, main-belt asteroids, and family members,
the physical disruptions of comets,
the Yarkovsky/YORP drift in semimajor axis,
chaotic diffusion in eccentricity/inclination,
or possible perturbations by the giant-planet migration.

Assuming a ``standard'' size-frequency distribution of primordial comets,
we predict the number of families with parent-body sizes $D_{\rm PB} \ge 200\,{\rm km}$
-- created during the LHB and subsequent $\simeq 4\,{\rm Gyr}$ of collisional evolution --
which seems consistent with observations. However, more than 100 asteroid families
with $D_{\rm PB} \ge 100\,{\rm km}$ should be created at the same time which are not observed.
This discrepancy can be nevertheless explained by the following processes:
  i)~asteroid families are efficiently destroyed by comminution (via collisional cascade),
 ii)~disruptions of comets below some critical perihelion distance ($q \lesssim 1.5\,{\rm AU}$) are common.

Given the freedom in the cometary-disruption law, we cannot provide stringent limits
on the cometary flux, but we can conclude that the observed distribution of asteroid families
does not contradict with a cometary LHB.}


\maketitle

%%%%%%%%%%%%%%%%%%%%%%%%%%%%%%%%%%%%%%%%%%%%%%%%%%%%%%%%%%%%%%%%%%%%%%%%

\section{Introduction}

The late heavy bombardment (LHB) is an important period in the history 
of the solar system. It is often defined as the process that made the huge 
but relatively young impact basins (a 300 km or larger diameter crater)
on the Moon like Imbrium and Orientale.
The sources and extent of the LHB, however, has been undergoing 
recent revisions. In the past, there were two end-member schools of 
thought describing the LHB. The first school argued that nearly all 
lunar basins, including the young ones, were made by impacting 
planetesimals left over from terrestrial planet formation
(Neukum et al. 2001, Hartmann et al. 2000, 2007; see Chapman et al. 2007 for a 
review). The second school argued that most lunar basins were made 
during a spike of impacts that took place near 3.9 Ga (e.g., Tera et al. 1974, Ryder et al. 2000).

Recent studies, however, suggest that a compromise scenario may be the best 
solution: the oldest basins were mainly made by leftover planetesimals, 
while the last 12--15 or so lunar basins were created by asteroids driven 
out of the primordial main belt by the effects of late giant-planet migration
(Tsiganis et al. 2005, Gomes et al. 2005, Minton \& Malhotra 2009,
Morbidelli et al. 2010, Marchi et al. 2012, Bottke et al. 2012).  
This would mean the LHB is limited in extent and does not encompass all 
lunar basins. If this view is correct, we can use studies of lunar and 
asteroid samples heated by impact events, together with dynamical 
modelling work, to suggest that the basin-forming portion of the LHB 
lasted from approximately 4.1--4.2 to 3.7--3.8 billion years ago on the Moon
(Bogard 1995, 2011, Swindle et al. 2009, Bottke et al. 2012, Norman \& Nemchin 2012).

The so-called `Nice model' provides a coherent explanation of the
origin of the LHB as an impact spike or rather a ``sawtooth'' (Morbidelli et al. 2012).
According to this model, the bombardment was triggered by a late dynamical orbital instability of
the giant planets, in turn driven by the gravitational interactions
between said planets and a massive transneptunian disk of
planetesimals (see Morbidelli 2010 for a review). In this scenario,
three projectile populations contributed to the LHB: the comets from
the original transneptunian disk (Gomes et al. 2005), the asteroids
from the main belt (Morbidelli et al. 2010) and those from a putative
extension of the main belt towards Mars, inwards of its current inner
edge (Bottke et al. 2012). The last could have been enough of a
source for the LHB, as recorded in the lunar crater record (Bottke et
al. 2012), while the asteroids from the current main belt boundaries
would have only been a minor contributor (Morbidelli et al. 2010).

The Nice model, however, predicts a very intense cometary bombardment
of which there seems to be no obvious traces on the Moon.  In fact, given the
expected total mass in the original transneptunian disk (Gomes et
al. 2005) and the size distribution of objects in this disk
(Morbidelli et al. 2009), the Nice model predicts that about
$5\times 10^4$ km-size comets should have hit the Moon during the LHB. This
would have formed 20\,km craters with a surface density of $1.7\times
10^{-3}$ craters per km$^2$. But the highest crater densities of 20\,km
craters on the lunar highlands is less than $2\times 10^{-4}$ (Strom
et al. 2005). This discrepancy might be explained by a gross
overestimate of the number of small bodies in the original
transneptunian disk in Morbidelli et al. (2009). However, all impact
clast analyses of samples associated to major LHB basins (Kring and
Cohen 2002, Tagle 2005) show that also the major projectiles were
not carbonaceous chondrites or similar primitive, comet-like objects.

The lack of evidence of a cometary bombardment of the Moon can be
considered as a fatal flaw in the Nice model. Curiously, however, in
the outer solar system we see evidence of the cometary flux predicted
by the Nice model. Such a flux is consistent with the number of impact
basins on Iapetus (Charnoz et al. 2009), with the number and the size
distribution of the irregular satellites of the giant planets
(Nesvorn\'y et al. 2007, Bottke et al. 2010) and of the Trojans of
Jupiter (Morbidelli et al. 2005), as well as with the capture of
D-type asteroids in the outer asteroid belt (Levison et al.,
2009). Moreover, the Nice model cometary flux is required to explain
the origin of the collisional break-up of the asteroid (153) Hilda in
the 3/2 resonance with Jupiter (located at $\simeq 4$ AU, i.e. beyond
the nominal outer border of the asteroid belt at $\simeq 3.2$ AU; Bro\v z
et al. 2011).

Missing signs of an intense cometary bombardment on the Moon and the
evidence for a large cometary flux in the outer solar system suggest
that the Nice model may be correct in its basic features, but
most comets disintegrated as they penetrated deep into the inner solar
system.

To support or reject this possibility, this paper focusses on the main
asteroid belt, looking for constraints on the flux of comets through
this region at the time of the LHB. In particular we focus on old
asteroid families, produced by the collisional break-up of large
asteroids, which may date back at the LHB time. We provide a census of
these families in Section~\ref{sec:families}.

In Section~\ref{sec:collisions_MB_alone}, we construct a collisional
model of the main belt population. We show that, on average, this
population alone could not have produced the observed number of
families with $D_{\rm PB}=200$--400\,km.
Instead, the required number of families with large parent bodies
is systematically produced if the asteroid belt was crossed by a large
number of comets during the LHB, as expected in the Nice model (see Section~\ref{sec:collisions_MB_comets}).
However, for any reasonable size distribution of the cometary population,
the same cometary flux that would produce the correct number
of families with $D_{\rm PB}=200$--400\,km would produce too many families
with $D_{\rm PB}\simeq 100$\,km relative to what is observed. Therefore, in
the subsequent sections we look for mechanisms that might prevent
detection of most of these families.

More specifically, in Sec.~\ref{sec:overlap} we discuss the possibility that families with
$D_{\rm PB}\simeq 100$\,km are so numerous that they cannot be identified
because they overlap with each other. In Sec.~\ref{sec:yarko_dispersal} we investigate their possible dispersal
below detectability due to the Yarkovsky effect and chaotic diffusion.
In Sec.~\ref{sec:comets_lifetime} we discuss the role of the physical lifetime of comets.
In Sec.~\ref{sec:jumping_jupiter} we analyse the dispersal of families
due to the changes in the orbits of the giant planets expected in the Nice model.
In Sec.~\ref{sec:comminution} we consider the subsequent collisional comminution
of the families. Of all investigated processes, the last one seems to be the most promising
for reducing the number of visible families with $D_{\rm PB}\simeq 100$\,km while
not affecting the detectability of old families with $D_{\rm PB}=200$--400\,km.

Finally, in Section~\ref{sec:pristine} we analyse a curious portion of
the main belt, located in a narrow semi-major axis zone bounded by the
5:2 and 7:3 resonances with Jupiter. This zone is severely deficient
in small asteroids compared to the other zones of the main belt. For
the reasons explained in the section, we think that this zone best
preserves the initial asteroid belt population, and therefore we call
it the ``pristine zone''. We checked the number of families in the
pristine zone, their sizes, and ages and we found that they are consistent
with the number expected in our model invoking a cometary bombardment
at the LHB time and a subsequent collisional
comminution and dispersion of the family members.

The conclusions follow in Section~\ref{sec:conclusions}.

%%%%%%%%%%%%%%%%%%%%%%%%%%%%%%%%%%%%%%%%%%%%%%%%%%%%%%%%%%%%%%%%%%%%%%%%

\section{A list of known families}\label{sec:families}

Although several lists of families exist in the literature (Zappal\'a et al. 1995, Nesvorn\'y et al. 2005, Parker et al. 2008, Nesvorn\'y 2010),
we are going to identify the families once again.
The reason is that we seek an {\em upper limit\/} for the number of {\em old families\/}
that may be significantly dispersed and depleted,
while the previous works often focussed on well-defined families.
Moreover, we need to calculate several {\em physical parameters\/} of the families
(such as the parent-body size, slopes of the size-frequency distribution,
a dynamical age estimate if not available in the literature) which are crucial for further modelling.
Last but not least, we use more precise {\em synthetic\/} proper elements from the AstDyS database
(Kne\v zevi\'c \& Milani 2003, version Aug 2010) instead of semi-analytic ones.

We employed a hierarchical clustering method (HCM, Zappal\'a et al. 1995) for the {\em initial\/} identification of families
in the proper element space $(a_{\rm p}, e_{\rm p}, \sin I_{\rm p})$,
but then we had to perform a lot of manual operations, because
i)~we had to select a reasonable cut-off velocity $v_{\rm cutoff}$,
usually such that the number of members~$N(v_{\rm cutoff})$ increases
relatively slowly with increasing~$v_{\rm cutoff}$.
ii)~The resulting family should also have a ``reasonable'' shape in the space of proper elements,
which should somehow correspond to the local dynamical features.%
\footnote{For example, the Eos family has a complicated but still reasonable shape,
since it is determined by several intersecting high-order mean-motion or secular resonances, see Vokrouhlick\'y et al. (2006).}
iii)~We checked taxonomic types (colour indices from the Sloan DSS MOC catalogue version 4, Parker et al. 2008),
which should be consistent among family members. We can recognise interlopers or overlapping families this way.
iv)~Finally, the size-frequency distribution should exhibit one or two well-defined slopes,
otherwise the cluster is considered uncertain.

\begin{figure*}
\centering
\begin{tabular}{cc}
families & background \\
\includegraphics[width=9.0cm]{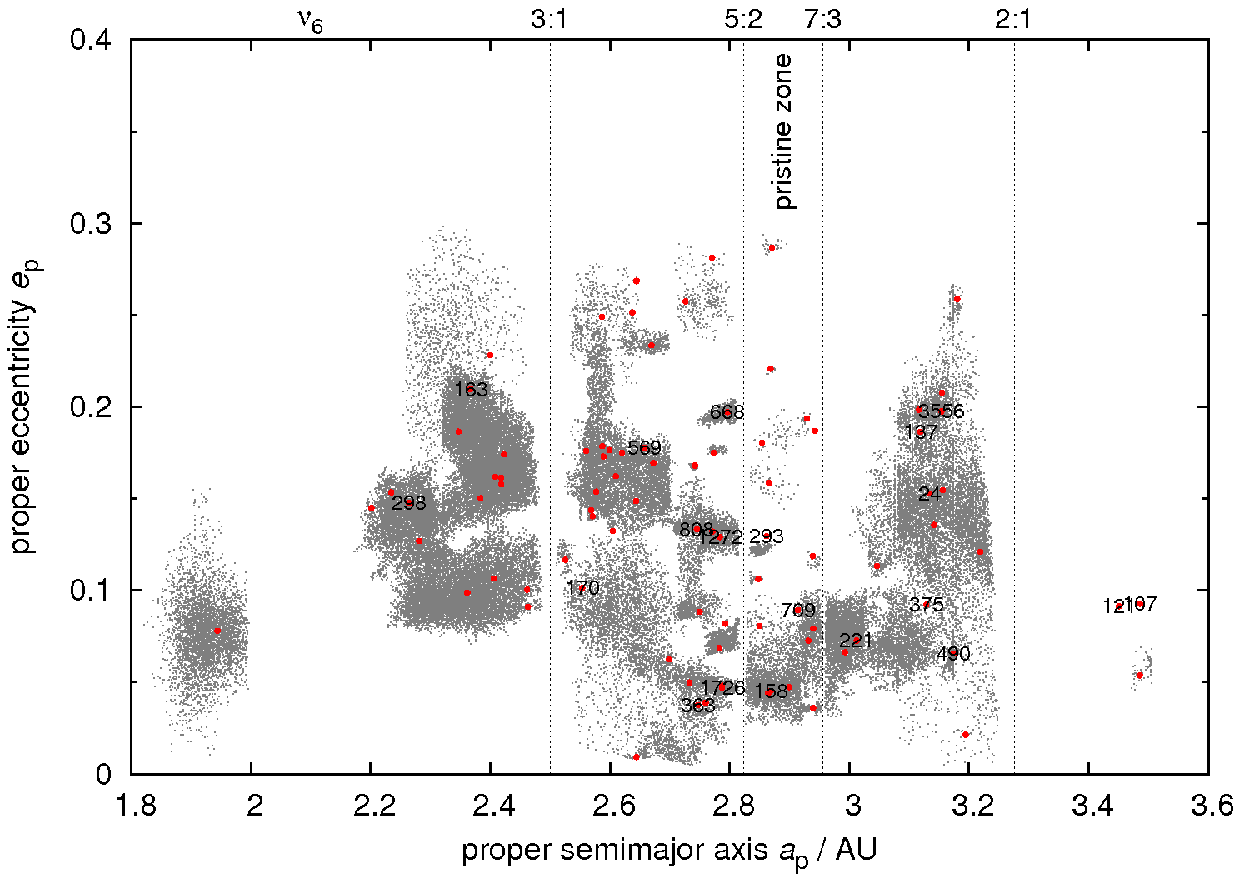} & \includegraphics[width=9.0cm]{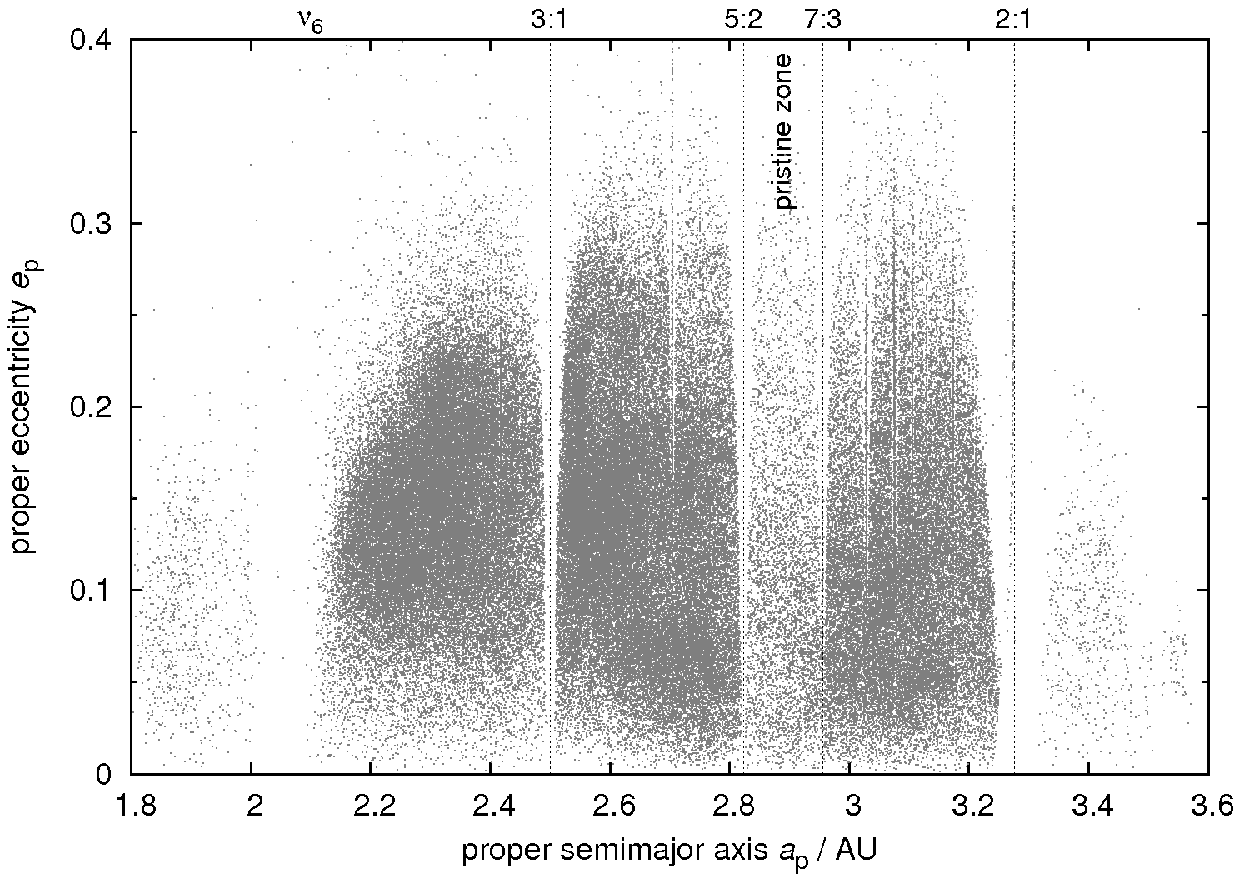} \\
\includegraphics[width=9.0cm]{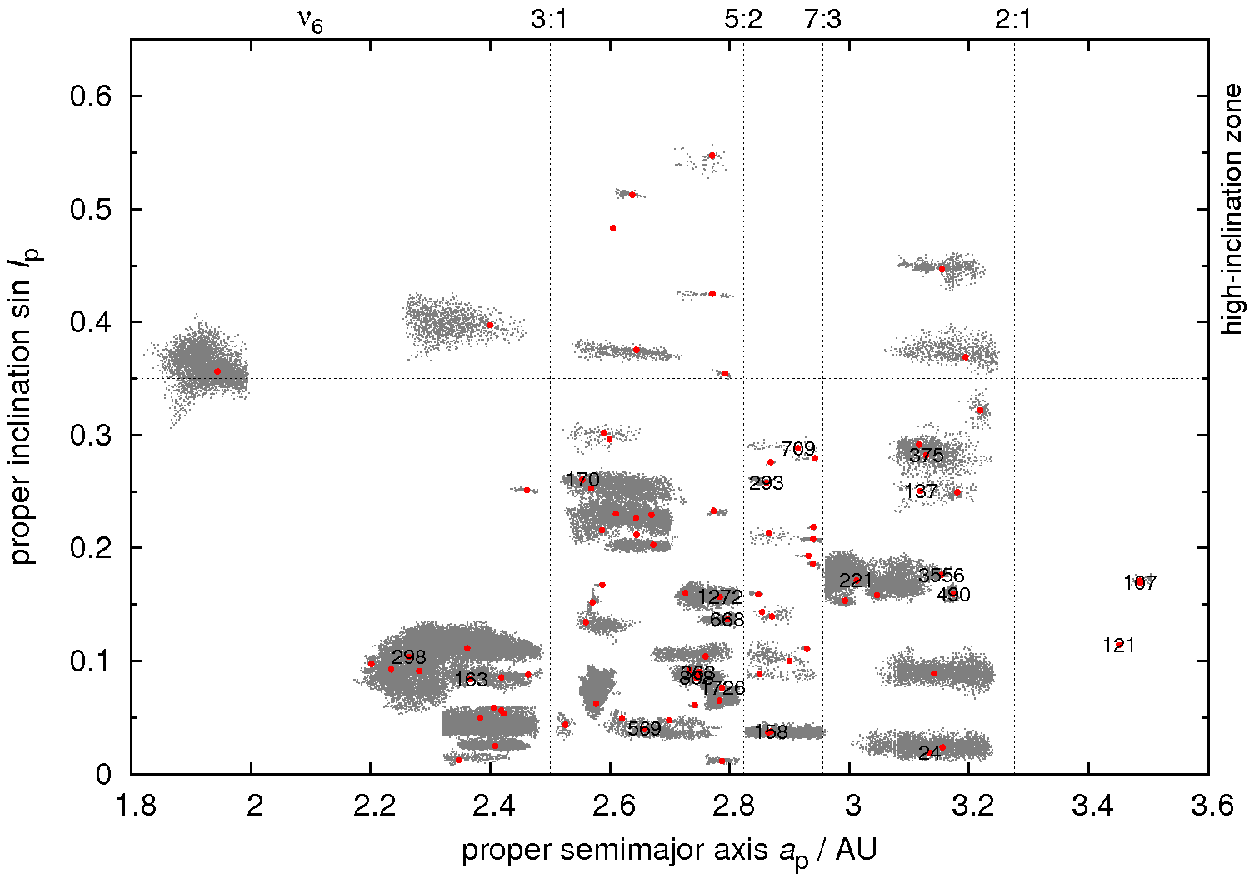} & \includegraphics[width=9.0cm]{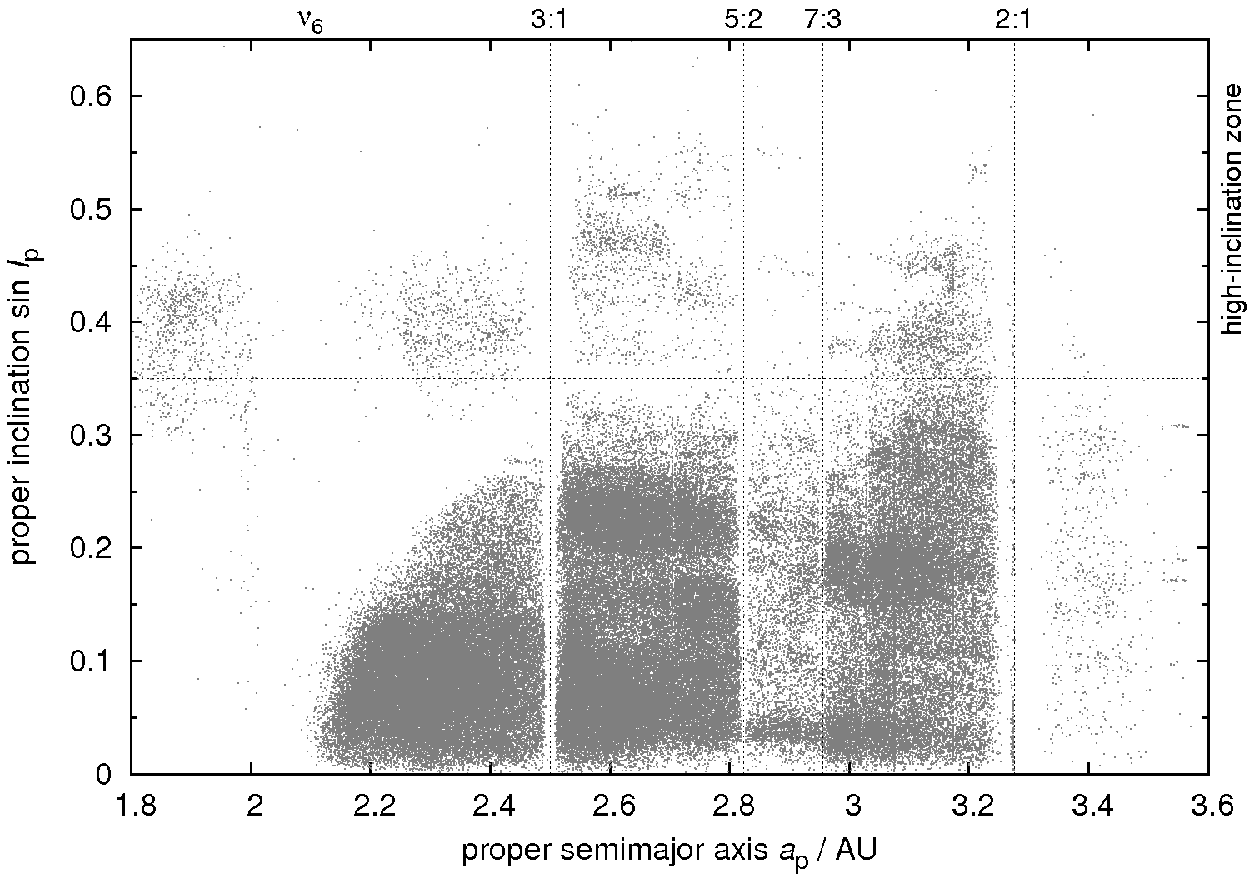} \\
\end{tabular}
\caption{Asteroids from the synthetic AstDyS catalogue plotted
in the proper semimajor axis~$a_{\rm p}$ vs proper eccentricity~$e_{\rm p}$ (top panels)
and $a_{\rm p}$ vs proper inclination $\sin I_{\rm p}$ planes (bottom panels).
We show the identified asteroid families (left panels)
with the positions of the largest members indicated by red symbols,
and also remaining background objects (right panels).
The labels correspond to designations of the asteroid families
that we focus on in this paper.
There are still some structures consisting of {\em small\/} objects
in the background population, visible only in the inclinations (bottom right panel).
These ``halos'' may arise for two reasons:
i)~a family has no sharp boundary and its transition to the background is smooth, or
ii)~there are bodies escaping from the families due to long-term dynamical evolution.
Nevertheless, we checked that these halo objects do not significantly affect our
estimates of parent-body sizes.}
\label{families_ae}
\end{figure*}

Our results are summarised in online Tables~\ref{FAMILIES_TABLE}--\ref{FAMILIES_TABLE3}
and the positions of families within the main belt are plotted in Figure~\ref{families_ae}.
Our list is ``optimistic'', so that even not very prominent families are included here.%
\footnote{On the other hand, we do not include all of the small and less-certain clumps
in a high-inclination region as listed by Novakovi\'c et al. (2011). We do not
focus on small or high-$I$ families in this paper.}

There are, however, several potential problems we are aware of:
\begin{enumerate}
\item There may be {\em inconsistencies\/} among different lists of families.
For example, sometimes a clump may be regarded as a single family or as two separate families.
This may be the case of:
Padua and Lydia,
Rafita and Cameron.

\item To identify families we used {\em synthetic\/}
proper elements, which are more precise than the semi-analytic
ones. Sometimes the families look more regular (e.g., Teutonia) or
more tightly clustered (Beagle) when we use the synthetic elements.
This very choice may, however, affect results substantially!
A clear example is the Teutonia family, which also contains the {\em big\/}
asteroid (5)~Astraea if the synthetic proper elements are
used, but {\em not\/} if the semi-analytic proper elements are used. 
This is due to the large differences between the semi-analytic and
synthetic proper elements of (5)~Astraea. Consequently, the 
physical properties of the two families differ considerably.
We believe that the family defined from the synthetic elements
is more reliable. 

\item Durda et al. (2007) often claim a {\em larger\/} size for the parent body
(e.g., Themis, Meliboea, Maria, Eos, Gefion),
because they try to match the SFD of larger bodies and the results of SPH experiments.
This way they also account for small bodies that existed at the time of the disruption,
but which do {\em not\/} exist today since they were lost due to collisional grinding
and the Yarkovsky effect. We prefer to use $D_{\rm Durda}$
instead of the value $D_{\rm PB}$ estimated from the currently observed SFD.
The geometric method of Tanga et al. (1999),
which uses the sum of the diameters of the first and third largest family members
as a first guess of the parent-body size, is essentially similar to our approach.

\end{enumerate}

A complete list of all families' members is available at our web site
{\tt http://sirrah.troja.mff.cuni.cz/\-\~{}mira/mp/fams/}, including
supporting figures.

%%%%%%%%%%%%%%%%%%%%%%%%%%%%%%%%%%%%%%%%%%%%%%%%%%%%%%%%%%%%%%%%%%%%%%%%

\newcommand{\tableheader}{
\hline\hline
\multicolumn{2}{c}{designation} & $v_{\rm cutoff}$ & $N$ & $p_V$ & tax. & $D_{\rm PB}$ & $D_{\rm Durda}$ & LR/PB & $v_{\rm esc}$ & $q_1$ & $q_2$ & age & notes, references \\
\multicolumn{2}{c}{} & m/s & & & & km & km & & m/s & & & Gyr & \\
\hline
}

\onltab{1}{
\begin{sidewaystable}[p]
\centering
\caption{%
A list of asteroid families and their physical parameters.
There are the following columns:
$v_{\rm cutoff}$~is the selected cut--off velocity for the hierarchical clustering,
$N$~the corresponding number of family members,
$p_V$~the adopted value of the geometric albedo for family members which do not have measured diameters
(from Tedesco et al. 2002 or Masiero et al. 2011, a letter 'w' indicates it was necessary to use the WISE data to obtain median/mean albedo),
taxonomic classification (according to the Sloan DSS MOC~4 colours, Parker et al. 2008),
$D_{\rm PB}$~parent body size, an additional 'c' letter indicates that we prolonged the SFD slope down to zero~$D$ (a typical uncertainty is 10\,\%),
$D_{\rm Durda}$~PB size inferred from SPH simulations (Durda et al. 2007), an exclamation mark denotes a significant mismatch with~$D_{\rm PB}$,
LR/PB~the ratio of the volumes of the largest remnant to the parent body (an uncertainty corresponds to the last figure, a range is given if both $D_{\rm PB}$ and $D_{\rm Durda}$ are known),
$v_{\rm esc}$~the escape velocity,
$q_1$~the slope of the SFD for larger~$D$,
$q_2$~the slope for smaller~$D$ (a typical uncertainty of the slopes is 0.2, if not indicated otherwise),
dynamical age including its uncertainty.}
\label{FAMILIES_TABLE}
\begin{tabular}{rlrrlllllll@{\kern3pt}l@{\kern3pt}lp{5.55cm}}
\tableheader
3 & Juno		& 50		& 449		& 0.250	& S	& 233	& ?		& 0.999		& 139		& $-4.9$	& $-3.2$	& $<$0.7	& cratering, Nesvorn\'y et al. (2005) \\
4 & Vesta		& 60		& 11169		& 0.351w& V	& 530	& 425!		& 0.995		& 314		& $-4.5$	& $-2.9$	& $1.0\pm0.25$	& cratering, Marchi et al. (2012) \\
8 & Flora		& 60		& 5284		& 0.304w& S	& 150c	& 160		& 0.81-0.68	& 88		& $-3.4$	& $-2.9$	& $1.0\pm0.5$	& cut by $\nu_6$ resonance, LL chondrites \\
10 & Hygiea		& 70		& 3122		& 0.055	& C,B	& 410	& 442		& 0.976-0.78	& 243		& $-4.2$	& $-3.2$	& $2.0\pm1.0$	& LHB? cratering \\
15 & Eunomia		& 50		& 2867		& 0.187	& S	& 259	& 292		& 0.958-0.66	& 153		& $-5.6$	& $-2.3$	& $2.5\pm0.5$	& LHB? Michel et al. (2002) \\
20 & Massalia		& 40		& 2980		& 0.215	& S	& 146	& 144		& 0.995		& 86		& $-5.0$	& $-3.0$	& $0.3\pm0.1$	& \\
24 & Themis		& 70		& 3581		& 0.066	& C	& 268c	& 380-430!	& 0.43-0.09	& 158		& $-2.7$	& $-2.4$	& $2.5\pm1.0$	& LHB? \\
44 & Nysa (Polana)	& 60		& 9957		& 0.278w& S	& 81c	& ?		& 0.65		& 48		& $-6.9$	& $-2.6$(0.5)	& $<$1.5	& overlaps with the Polana family \\ % !!!
46 & Hestia		& 65		& 95		& 0.053	& S	& 124	& 153		& 0.992-0.53	& 74		& $-3.3$	& $-2.0$	& $<$0.2	& cratering, close to J3/1 resonance \\
87 & Sylvia		& 110		& 71		& 0.045	& C/X	& 261	& 272		& 0.994-0.88	& 154		& $-5.2$	& $-2.4$	& 1.0-3.8	& LHB? cratering, Vokrouhlick\'y et al. (2010) \\
128 & Nemesis		& 60		& 654		& 0.052	& C	& 189	& 197		& 0.987-0.87	& 112		& $-3.4$	& $-3.3$	& $0.2\pm0.1$	& \\
137 & Meliboea		& 95		& 199		& 0.054	& C	& 174c	& 240-290!	& 0.59-0.20	& 102		& $-1.9$	& $-1.8$	& $<$3.0	& old? \\ % !!!
142 & Polana (Nysa)	& 60		& 3443		& 0.055w& C	& 75	& ?		& 0.42		& 45		& $-7.0$	& $-3.6$	& $<$1.5	& overlaps with Nysa \\ % !!!
145 & Adeona		& 50		& 1161		& 0.065	& C	& 171c	& 185		& 0.69-0.54	& 101		& $-5.2$	& $-2.8$	& $0.7\pm0.5$	& cut by J5/2 resonance \\
158 & Koronis		& 50		& 4225		& 0.147	& S	& 122c	& 170-180	& 0.024-0.009	& 68		& $-3.6(0.3)$	& $-2.3$	& $2.5\pm1.0$	& LHB? \\
163 & Erigone		& 60		& 1059		& 0.056	& C/X	& 79	& 114		& 0.79-0.26	& 46		& ?		& $-3.6$	& $0.3\pm0.2$	& \\
170 & Maria		& 80		& 3094		& 0.249w& S	& 107c	& 120-130	& 0.070-0.048	& 63		& $-2.5(0.3)$	& $-2.8$	& $3.0\pm1.0$	& LHB? \\
221 & Eos		& 50		& 5976		& 0.130 & K	& 208c	& 381!		& 0.13-0.020	& 123		& $-3.5$	& $-2.1$	& $1.3\pm0.2$	& \\
283 & Emma		& 75		& 345		& 0.050	& -	& 152	& 185		& 0.92-0.51	& 90		& ?		& $-3.2$	& $<$1.0	& satellite \\
293 & Brasilia		& 60		& 282		& 0.175w& C/X	& 34	& ?		& 0.020		& 20		& $-1.4(0.5)$	& $-3.7$	& $0.05\pm0.04$	& (293) is interloper \\
363 & Padua (Lydia)	& 50		& 596		& 0.097	& C/X	& 76	& 106		& 0.045-0.017	& 45		& $-1.8$	& $-3.0$	& $0.3\pm0.2$	& \\
396 & Aeolia		& 20		& 124		& 0.171	& C/X	& 35	& 39		& 0.966-0.70	& 20		& ?		& $-4.3$	& $<$0.1	& cratering \\
410 & Chloris		& 90		& 259		& 0.057	& C	& 126c	& 154		& 0.952-0.52	& 74		& ?		& $-2.1$	& $0.7\pm0.4$	& \\
490 & Veritas		& -		& -		& -	& C,P,D	& -	& 100-177	& -		& -		& -		& -		& $0.0083\pm0.0005$	& (490) is likely interloper (Michel et al. 2011) \\
569 & Misa		& 70		& 543		& 0.031	& C	& 88c	& 117		& 0.58-0.25	& 52		& $-3.9$	& $-2.3$	& $0.5\pm0.2$	& \\
606 & Brangane		& 30		& 81		& 0.102	& S	& 37	& 46		& 0.92-0.48	& 22		& ?		& $-3.8$	& $0.05\pm0.04$	& \\
668 & Dora		& 50		& 837		& 0.054	& C	& 85	& 165!		& 0.031-0.004	& 50		& $-4.2$	& $-1.9$	& $0.5\pm0.2$	& \\
808 & Merxia		& 50		& 549		& 0.227	& S	& 37	& 121!		& 0.66-0.018	& 22		& $-2.7$	& $-3.4$	& $0.3\pm0.2$	& \\
832 & Karin		& -		& -		& -	& S	& -	& 40		& -		& -		& -		& -		& $0.0058\pm0.0002$	& \\
845 & Naema		& 30		& 173		& 0.081	& C	& 77c	& 81		& 0.35-0.30	& 46		& $-5.2$	& $-2.9$	& $0.1\pm0.05$	& \\
847 & Agnia		& 40		& 1077		& 0.177	& S	& 39	& 61		& 0.38-0.10	& 23		& $-2.8$	& $-3.1$	& $0.2\pm0.1$	& \\
1128 & Astrid		& 50		& 265		& 0.079	& C	& 43c	& ?		& 0.52		& 25		& $-1.7$	& $-2.6$	& $0.1\pm0.05$	& \\
1272 & Gefion		& 60		& 19477		& 0.20	& S	& 74c	& 100-150!	& 0.001-0.004	& 60		& $-4.3$	& $-2.5$	& $0.48\pm0.05$	& Nesvorn\'y et al. (2009), L chondrites \\
1400 & Tirela		& 80		& 1001		& 0.070	& S	& 86	& -		& 0.12		& 86		& $-4.2$	& $-3.4$	& $<$1.0	& \\
1658 & Innes		& 70		& 621		& 0.246w& S	& 27	& ?		& 0.14		& 16		& $-4.9$	& $-3.5$	& $<$0.7	& (1644) Rafita is interloper \\
1726 & Hoffmeister	& 40		& 822		& 0.035	& C	& 93c	& 134		& 0.022-0.007	& 55		& $-4.5$	& $-2.7$	& $0.3\pm0.2$	& \\
3556 & Lixiaohua	& 60		& 439		& 0.044w& C/X	& 62	& 220!		& 0.029-0.001	& 35		& $-6.1$	& $-3.3$	& $0.15\pm0.05$	& Novakovi\'c et al. (2010) \\
3815 & Konig		& 60		& 177		& 0.044	& C	& 33	& ?		& 0.32		& 20		& ?		& $-3.0$	& $<$0.1	& (1639) Bower is interloper \\
4652 & Iannini		& -		& -		& -	& S	& -	& -		& -		& -		& -		& -		& $0.005\pm0.005$	&  \\
9506 & Telramund	& 40		& 146		& 0.217w& S	& 22	& -		& 0.05		& 13		& $-3.9$	& $-3.7$	& $<$0.5	& \\
18405 & 1993 FY$_{12}$	& 50		& 44		& 0.171w& C/X	& 15	& -		& 0.23		& 15		& $-2.4$	& $-2.4$	& $<$0.2	& cut by J5/2 resonance \\
\hline
\end{tabular}
%\vspace{-18cm}
\end{sidewaystable}
\hbox{}\newpage

}

\onltab{2}{
\begin{sidewaystable}[!p]
\caption{Continuation of Table~\ref{FAMILIES_TABLE}.}
\label{FAMILIES_TABLE2}
\centering
\begin{tabular}{rlrrllll@{\kern3pt}lllllp{7.0cm}}
\tableheader

158 & Koronis$_{(2)}$	& -		& -		& -	& S	& 35	& -		& -		& -		& -		& -		& $0.015\pm0.005$	& cratering, Molnar \& Haegert (2009) \\
298 & Baptistina	& 50		& 1249		& 0.160w& C/X	& 35c	& -		& 0.17		& 21		& $-3.6$	& $-2.4$	& $<$0.3	& part of the Flora family\\
434 & Hungaria		& 200		& 4598		& 0.35	& E	& 25	& -		& 0.15		& 15		& $-5.9$	& $-3.1$	& $0.5\pm0.2$	& Warner et al. (2010) \\
627 & Charis		& 80		& 235		& 0.081	& S	& $>$60	& -		& 0.53		& 35		& ?		& $-3.4$	& $<$1.0	& \\
778 & Theobalda		& 85		& 154		& 0.060	& C	& 97c	& -		& 0.29		& 57		& ?		& $-2.9$	& $0.007\pm0.002$	& cratering, Novakovi\'c (2010) \\
\noalign{\medskip}
302 & Clarissa		& 30		& 75		& 0.054	& C	& 39	& -		& 0.96		& 23		& ?		& $-3.1$	& $<$0.1	& cratering, Nesvorn\'y (2010) \\
656 & Beagle		& 24		& 63		& 0.089	& C	& 64	& -		& 0.56		& 38		& $-1.3$	& $-1.4$	& $<$0.2	& \\
752 & Sulamitis		& 60		& 191		& 0.042	& C	& 65	& -		& 0.83		& 39		& $-6.5$	& $-2.3$	& $<$0.4	& \\
1189 & Terentia		& 50		& 18		& 0.070	& C	& 56	& -		& 0.990		& 33		& ?		& $-2.6$?	& $<$0.2	& cratering \\
1892 & Lucienne		& 100		& 57		& 0.223w& S	& 14	& -		& 0.71		& 8		& ?		& $-4.4$	& $<$0.3	& \\
7353 & Kazvia		& 50		& 23		& 0.206w& S	& 16	& -		& 0.57		& 8		& ?		& $-1.8$	& $<$0.1	& \\
10811 & Lau		& 100		& 15		& 0.273w& S	& 11	& -		& 0.77		& 5		& ?		& $-2.8$	& $<$0.1	& \\
18466 & 1995 SU$_{37}$	& 40		& 71		& 0.241w& S	& 14	& -		& 0.045		& 7		& ?		& $-5.0$	& $<$0.3	& \\
\noalign{\medskip}
1270 & Datura		& -		& -		& -	& S	& -	& -		& -		& -		& -		& -		& 0.00045-0.00060	& identified in osculating-element space, \\
14627 & Emilkowalski	& -		& -		& -	& C/X	& -	& -		& -		& -		& -		& -		& 0.00019-0.00025	& Nesvorn\'y \& Vokrouhlick\'y (2006) \\
16598 & 1992 YC$_2$	& -		& -		& -	& S	& -	& -		& -		& -		& -		& -		& 0.00005-0.00025	& \\
21509 & Lucascavin	& -		& -		& -	& S	& -	& -		& -		& -		& -		& -		& 0.0003-0.0008	& \\
2384 & Schulhof		& -		& -		& -	& S	& -	& -		& -		& -		& -		& -		& 0.0007-0.0009	& Vokrouhlick\'y \& Nesvorn\'y (2011) \\
\noalign{\medskip}
27 & Euterpe		& 70		& 268		& 0.260w& S	& 118c	& -		& 0.998		& 70		& $-2.9$	& $-2.2$	& $<$1.0	& cratering, Parker et al. (2008) \\ % !!!
375 & Ursula		& 80		& 777		& 0.057w& C	& 203c	& 240-280	& 0.71-0.43	& 120		& $-4.1$	& $-2.3$	& $<$3.5	& old? \\ % !!!
1044 & Teutonia		& 50		& 1950		& 0.343 & S	& 27-120& -		& 0.17-0.98	& 16-71		& $-3.5$	& $-3.9$	& $<$0.5	& depends on (5) Astraea membership \\
1296 & Andree		& 60		& 401		& 0.290w& S	& 17-74	& -		& 0.010-0.95	& 10-43		& ?		& $-2.9(0.5)$	& $<$1.0	& depends on (79) Eurynome membership \\ % !!!
2007 & McCuskey		& 34		& 236		& 0.06  & C	& 29	& -		& 0.41		& 17		& ?		& $-5.6$	& $<$0.5	& overlaps with Nysa/Polana \\
2085 & Henan		& 54		& 946		& 0.200w& S	& 27	& -		& 0.13		& 16		& $-4.2$	& $-3.2$	& $<$1.0	& \\ % !!!
2262 & Mitidika		& 83		& 410		& 0.064w& C	& 49-79c& -		& 0.037-0.81	& 26-46		& $-4.5$	& $-2.2$	& $<$1.0	& depends on (785) Zwetana membership,\hfil\break (2262) is interloper, overlaps with Juno \\ % !!!
\noalign{\medskip}
2 & Pallas		& 200		& 64		& 0.163	& B	& 498c	& -		& 0.9996	& 295		& ?		& $-2.2$	& $<$0.5	& high-$I$, Carruba (2010) \\
25 & Phocaea		& 160		& 1370		& 0.22	& S	& 92	& -		& 0.54		& 55		& $-3.1$	& $-2.4$	& $<$2.2	& old? high-$I/e$, cut by $\nu_6$ resonance, Carruba (2009) \\ % !!!
148 & Gallia		& 150		& 57		& 0.169	& S	& 98	& -		& 0.058		& 58		& ?		& $-3.6$	& $<$0.45	& high-$I$ \\
480 & Hansa		& 150		& 651		& 0.256	& S	& 60	& -		& 0.83		& 35		& $-4.9$	& $-3.2$	& $<$1.6	& high-$I$ \\
686 & Gersuind		& 130		& 178		& 0.146	& S	& 52c	& -		& 0.48		& 40		& ?		& $-2.7$	& $<$0.8	& high-$I$, Gil-Hutton (2006) \\
945 & Barcelona		& 110		& 129		& 0.248	& S	& 28	& -		& 0.77		& 16		& ?		& $-3.5$	& $<$0.35	& high-$I$, Foglia \& Masi (2004) \\
1222 & Tina		& 110		& 37		& 0.338	& S	& 21	& -		& 0.94		& 12		& ?		& $-4.1$	& $<$0.15	& high-$I$ \\
4203 & Brucato		& -		& -		& -	& -	& -	& -		& -		& -		& -		& -		& $<$1.3	& in freq. space \\
\noalign{\medskip}                                                                                                                                                                                                                                                           
31 & Euphrosyne		& 100		& 851		& 0.056	& C	& 259	& -		& 0.97		& 153		& $-4.9$	& $-3.9$	& $<$1.5	& cratering, high-$I$, Foglia \& Massi (2004)  \\ % !!!
702 & Alauda		& 120		& 791		& 0.070	& B	& 218c	& 290-330!	& 0.025		& 129		& $-3.9$	& $-2.4$	& $<$3.5	& old? high-$I$, cut by J2/1 resonance, satellite\hfil\break (Margot \& Rojo 2007) \\
\noalign{\medskip}
107 & Camilla		& ?		& ?		& 0.054	& -	& $>$226& ?		& ?		& ?		& ?		& ?		& 3.8?		& LHB? Cybele region, non-existent today, \\
121 & Hermione		& ?		& ?		& 0.058	& -	& $>$209& ?		& ?		& ?		& ?		& ?		& 3.8?		& LHB? Vokrouhlick\'y et al. (2010) \\

\hline
\end{tabular}
\vspace{-18cm}
\end{sidewaystable}

}

\onltab{3}{
\begin{sidewaystable}[!p]
\caption{Continuation of Table~\ref{FAMILIES_TABLE}.}
\label{FAMILIES_TABLE3}
\centering
\begin{tabular}{rlrrlllllllllp{7.6cm}}
\tableheader

1303 & Luthera		& 100		& 142		& 0.043	& X	& 92	& -		& 0.81		& 54		& $-3.9$	& $-2.7$	& $<$0.5	& above (375) Ursula \\
1547 & Nele		& 20		& 57		& 0.311w& X	& 19	& -		& 0.85		& 11		& ?		& $-2.8(0.3)$	& $<$0.04	& close to (3) Juno \\
2732 & Witt		& 60		& 985		& 0.260w& S	& 25	& -		& 0.082		& 15		& $-4.0(0.3)$	& $-3.8$	& $<$1.0	& only part with $\sin I > 0.099$, above (363) Padua \\
\noalign{\medskip}                                                                                                                                                    
81 & Terpsichore	& 120		& 70		& 0.052	& C	& 119	& -		& 0.993		& 71		& ?		& $-4.4$	& $<$0.5	& cratering, less-certain families in the ``pristine zone'' \\
709 & Fringilla		& 140		& 60		& 0.047	& X	& 99c	& 130-140	& 0.93-0.41	& 59		& $-6.2$	& $-1.7$	& $<$2.5	& old? \\ % !!!
918 & Itha		& 140		& 63		& 0.23	& S	& 38	& -		& 0.16		& 22		& $-2.7$	& $-1.5$	& $<$1.5	& shallow SFD \\
5567 & Durisen		& 100		& 18		& 0.044w& X	& 42	& -		& 0.89		& 25		& ?		& $-1.7$	& $<$0.5	& shallow SFD \\
5614 & Yakovlev		& 100		& 34		& 0.05	& C	& 22	& -		& 0.28		& 13		& ?		& $-3.2$	& $<$0.2	& \\
12573 & 1999 NJ$_{53}$	& 40		& 13		& 0.190w& C	& 15	& -		& 0.13		& 9		& ?		& $-2.0(0.5)$	& $<$0.6	& incomplete SFD \\
15454 & 1998 YB$_{3}$	& 50		& 14		& 0.054w& C	& 21	& -		& 0.41		& 13		& ?		& $-1.6(0.3)$	& $<$0.5	& shallow SFD \\
15477 & 1999 CG$_{1}$	& 110		& 144		& 0.098w& S	& 25	& -		& 0.065		& 14		& ?		& $-4.6(0.5)$	& $<$1.5	& \\ % !!!
36256 & 1999 XT$_{17}$	& 60		& 30		& 0.210w& S	& 17	& -		& 0.037		& 10		& ?		& $-1.4(0.5)$	& $<$0.3	& shallow SFD \\

\hline
\end{tabular}
\vspace{-18cm}
\end{sidewaystable}

\hbox{}\newpage
}

%%%%%%%%%%%%%%%%%%%%%%%%%%%%%%%%%%%%%%%%%%%%%%%%%%%%%%%%%%%%%%%%%%%%%%%%

\subsection{A definition of the production function}

\begin{figure}
\centering
\includegraphics[width=8.0cm]{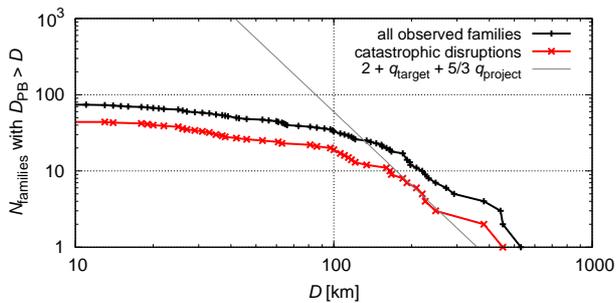}
\caption{A production function (i.e. the cumulative number~$N({>}D)$ of families with parent-body size~$D_{\rm PB}$ larger than~$D$)
for all observed families (black) and families corresponding to catastrophic disruptions (red),
i.e. with largest remnant/parent body mass ratio lower than 0.5. We also plot a theoretical slope according to Eq.~(\ref{eq:slope}),
assuming $q_{\rm target} = -3.2$ and $q_{\rm project} = -1.2$,
which correspond to the slopes of the main belt population
in the range $D = 100$--200\,km and $D = 15$--60\,km, respectively.}
\label{N_families}
\end{figure}

To compare observed families to simulations, we define a ``production function''
as the cumulative number~$N({>}D)$ of families with parent-body size~$D_{\rm PB}$ larger than a given~$D$.
The observed production function is shown in Figure~\ref{N_families},
and it is worth noting that it is very shallow.
The number of families with $D_{\rm PB} \simeq 100\,{\rm km}$
is comparable to the number of families in the $D_{\rm PB} = 200$--400\,km range.

It is important to note that the observed production function
is likely to be affected by biases (the family sample may not be complete, especially below $D_{\rm PB} \lesssim 100\,{\rm km}$)
and also by long-term collisional/dynamical evolution
which may prevent a detection of old comminutioned/dispersed families today
(Marzari et al. 1999).

From the theoretical point of view, the slope~$q$ of the production function $N({>}D) \propto D^q$
should correspond to the cumulative slopes of the size-frequency distributions
of the target and projectile populations. It is easy to show%
\footnote{Assuming that the strength is approximately $Q^\star_D \propto D^2$ in the gravity regime,
the necessary projectile size $d \propto (Q^\star_D)^{1/3} D$ (Bottke et al. 2005),
and the number of disruptions $n \propto D^2 D^{q_{\rm target}} d^{q_{\rm project}}$.}
that the relation is
\begin{equation}
q = 2 + q_{\rm target} + {5\over 3}q_{\rm project}\,.\label{eq:slope}
\end{equation}
Of course, real populations may have complicated SFDs, with different slopes in different ranges.
Nevertheless, any populations that have a steep SFD (e.g. $q_{\rm target} = q_{\rm project} = -2.5$)
would inevitably produce a steep production function ($q \doteq -4.7$).

In the following analysis, we drop cratering events
and discuss catastrophic disruptions only, i.e. families which have
largest remnant/parent body mass ratio less than 0.5.
The reason is that the same criterion ${\rm LR}/{\rm PB} < 0.5$ is used in collisional models.
Moreover, cratering events were not yet systematically explored by SPH simulations
due to insufficient resolution (Durda et al. 2007).

%%%%%%%%%%%%%%%%%%%%%%%%%%%%%%%%%%%%%%%%%%%%%%%%%%%%%%%%%%%%%%%%%%%%%%%%

\subsection{Methods for family age determination}\label{sec:ages}

If there is no previous estimate of the age of a family,
we used one of the following three dynamical methods to determine it:
  i)~a~simple $(a_{\rm p}, H)$ analysis as in Nesvorn\'y et al. (2005);
 ii)~a~$C$-parameter distribution fitting as introduced by Vokrouhlick\'y et al. (2006);
iii)~a~full N-body simulation described e.g. in Bro\v z et al. (2011).

\begin{figure}
\centering
\includegraphics[width=9.0cm]{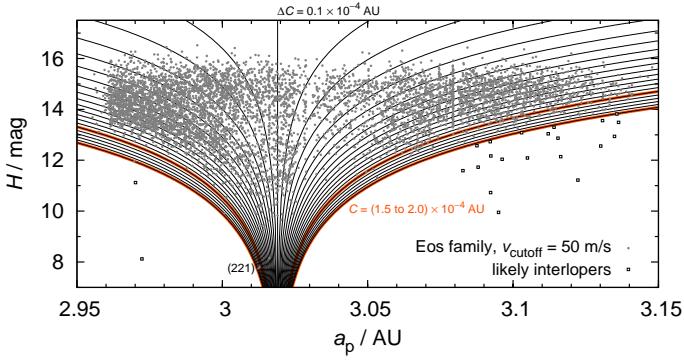}
\caption{An example of the Eos asteroid family, shown on the
proper semimajor axis~$a_{\rm p}$ vs absolute magnitude~$H$ plot.
We also plot curves defined by equation~(\ref{C_param})
and parameters~$a_{\rm c} = 3.019\,{\rm AU}$,
$C = 1.5\hbox{ to }2.0\times 10^{-4}\,{\rm AU}$,
which is related to the upper limit of the dynamical age of the family.}
\label{221_Eos_aH_CPARAM}
\end{figure}

In the first approach, we assume {\em zero\/} initial velocities,
and the current extent of the family is explained by the size-dependent Yarkovsky semimajor axis drift.
This way we can obtain only an {\em upper limit\/} for the dynamical age, of course.
We show an example for the Eos family in Figure~\ref{221_Eos_aH_CPARAM}.
The extent of the family in the proper semimajor axis vs the absolute magnitude
$(a_{\rm p}, H)$ plane can be described by the parametric relation
\begin{equation}
0.2 H = \log_{10} {|a_{\rm p} - a_{\rm c}| \over C}\,,\label{C_param}
\end{equation}
where $a_{\rm c}$ denotes the centre of the family, and $C$~is the parameter.
Such relation can be naturally expected when the semimajor-axis drift rate
is inversely proportional to the size, ${{\rm d}a/{\rm d}t} \propto 1/D$,
and the size is related to the absolute magnitude via the Pogson equation
$H = -2.5\log_{10} (p_V D^2/D_0^2)$, where $D_0$ denotes the reference diameter
and $p_V$ the geometric albedo (see Vokrouhlick\'y et al. 2006 for a detailed discussion).
The limiting value, for which all Eos family members (except interlopers)
are {\em above\/} the corresponding curve, is $C = 1.5\hbox{ to }2.0\times 10^{-4}\,{\rm AU}$.
Assuming reasonable thermal parameters (summarised in Table~\ref{tab:yarko}),
we calculate the expected Yarkovsky drift rates ${{\rm d}a/{\rm d}t}$ (using the theory from Bro\v z 2006)
and consequently can determine the age to be $t < 1.5\hbox{ to }2.0\,{\rm Gyr}$.

The second method uses a histogram~$N(C, C+\Delta C)$ of the number
of asteroids with respect to the~$C$ parameter defined above,
which is fitted by a dynamical model of the initial velocity field
and the Yarkovsky/YORP evolution. This enables us to determine the {\em lower limit\/}
for the age too (so the resulting age estimate is $t = 1.3^{+0.15}_{-0.2}\,{\rm Gyr}$ for the Eos family).

In the third case, we start an N-body simulation using a modified SWIFT
integrator (Levison and Duncan 1994), with the Yarkovsky/YORP acceleration included,
and evolve a synthetic family up to 4\,Gyr. We try to match the shape of the observed
family in all three proper orbital elements~$(a_{\rm p}, e_{\rm p}, \sin I_{\rm p})$.
In principle, this method may provide a somewhat independent estimate
of the age. For example, there is a `halo' of asteroids in the surroundings
of the nominal Eos family, which are of the same taxonomic type~K,
and we may fit the ratio $N_{\rm halo}/N_{\rm core}$ of the number of objects
in the `halo' and in the family `core' (Bro\v z et al., in preparation).

The major source of uncertainty in all methods are unknown bulk densities
of asteroids (although we use the most likely values for the S or C/X taxonomic classes, Carry 2012).
The age scales approximately as $t \propto \rho_{\rm bulk}$.
Nevertheless, we are still able to distinguish families that are young
from those that are old, because the allowed ranges of densities
for S-types ($2\hbox{ to }3\,{\rm g}/{\rm cm}^3$)
and C/X-types ($1\hbox{ to }2\,{\rm g}/{\rm cm}^3$) are limited (Carry 2012)
and so are the allowed ages of families.

\begin{table}
\caption{Nominal thermal parameters for S and C/X taxonomic types of asteroids:
$\rho_{\rm bulk}$~denotes the bulk density,
$\rho_{\rm surf}$~the surface density,
$K$~the thermal conductivity,
$C_{\rm th}$~the specific thermal capacity,
$A_{\rm Bond}$~the Bond albedo and
$\epsilon$~the infrared emissivity.}
\label{tab:yarko}
\begin{tabular}{ccccccc}
\hline
\hline
type & $\rho_{\rm bulk}$ & $\rho_{\rm surf}$ & $K$ & $C_{\rm th}$ & $A_{\rm Bond}$ & $\epsilon$ \\
     & (${\rm kg}/{\rm m}^3$) & (${\rm kg}/{\rm m}^3$) & (${\rm W}/{\rm m}/{\rm K}$) & (${\rm J}/{\rm kg}/{\rm K}$) & & \\
\hline
S   & 2500 & 1500 & 0.001 & 680 & 0.1  & 0.9 \\
C/X & 1300 & 1300 & 0.01  & 680 & 0.02 & 0.9 \\
\hline
\end{tabular}
\end{table}

%%%%%%%%%%%%%%%%%%%%%%%%%%%%%%%%%%%%%%%%%%%%%%%%%%%%%%%%%%%%%%%%%%%%%%%%

\subsection{Which families can be of LHB origin?}\label{sec:fams_LHB}

\begin{figure}
\centering
\includegraphics[width=9.0cm]{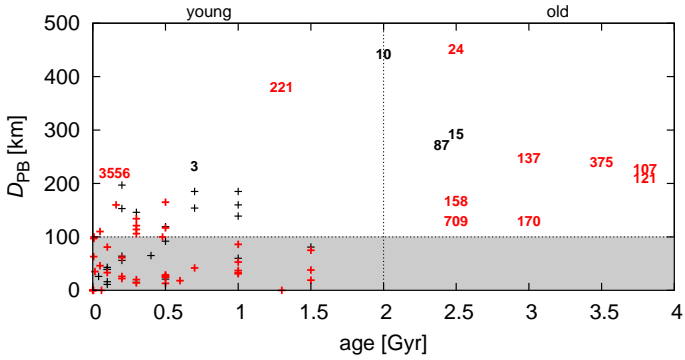}
\caption{The relation between dynamical ages of families and the sizes of their parent bodies.
Red labels correspond to catastrophic disruptions, while cratering
events are labelled in black. Some of the families are denoted by the designation
of the largest member. The uncertainties of both parameters are listed in
Tables~\ref{FAMILIES_TABLE}--\ref{FAMILIES_TABLE3} (we do not include overlapping
error bars here for clarity).}
\label{families_age}
\end{figure}

The ages of the observed families and their parent-body sizes are shown in Figure~\ref{families_age}.
Because the ages are generally very uncertain, we consider that any family whose nominal age is older than 2\,Gyr
is potentially a family formed ${\sim}4$\,Gyr ago, i.e. at the LHB time.
If we compare the number of ``young'' (${<}2\,{\rm Gyr}$) and old families (${>}2\,{\rm Gyr}$)
with $D_{\rm PB} = 200$--400\,km, we cannot see a significant over-abundance of old family formation events.
On the other hand, we almost do not find any small old families.

Only 12~families from the whole list may be {\em possibly\/} dated back
to the late heavy bombardment, because their dynamical ages approach $\sim 3.8\,{\rm Gyr}$
(including the relatively large uncertainties; see Table~\ref{tab:LHB_families},
which is an excerpt from Tables~\ref{FAMILIES_TABLE}--\ref{FAMILIES_TABLE3}).

If we drop cratering events and the families of Camilla and Hermione,
which do not exist any more today
(their existence was inferred from the satellite systems, Vokrouhlick\'y et al. 2010),
we end up with {\em only\/} five families
created by catastrophic disruptions that may potentially date from
the LHB time (i.e. their nominal age is more than 2\,Gy).
As we shall see in Section~\ref{sec:collisions_MB_comets}, this is an unexpectedly low number.

Moreover, it is really intriguing that most ``possibly-LHB'' families are larger
than $D_{\rm PB} \simeq 200\,{\rm km}$. It seems that old families with
$D_{\rm PB} \simeq 100\,{\rm km}$ are missing in the observed sample.
This is an important aspect that we have to explain, because it contradicts
our expectation of a steep production function.

\begin{table}
\centering
\caption{Old families with ages possibly approaching the LHB. They are sorted according
to the parent body size, where $D_{\rm Durda}$ determined by the Durda et al. (2007) method
is preferred to the estimate~$D_{\rm PB}$ inferred from the observed SFD.
An additional `c' letter indicates that we extrapolated the SFD down to $D = 0\,{\rm km}$
to account for small (unobserved) asteroids,
an exclamation mark denotes a significant mismatch between $D_{\rm PB}$ and $D_{\rm Durda}$.}
\label{tab:LHB_families}
\begin{tabular}{rlllll}
\hline
\hline
\multicolumn{2}{c}{designation} & $D_{\rm PB}$ & $D_{\rm Durda}$ & note \\
& & (km) & (km) & \\
\hline
24 & Themis		& 209c	& 380--430!	&  \\
10 & Hygiea		& 410	& 442		& cratering \\
15 & Eunomia		& 259	& 292		& cratering \\
702 & Alauda		& 218c	& 290--330!	& high-$I$ \\
87 & Sylvia		& 261	& 272		& cratering \\
137 & Meliboea		& 174c	& 240--290!	&  \\
375 & Ursula		& 198	& 240--280	& cratering \\
107 & Camilla		& $>$226& -		& non-existent \\
121 & Hermione		& $>$209& -		& non-existent \\
158 & Koronis		& 122c	& 170--180	&  \\
709 & Fringilla		& 99c	& 130--140	& cratering \\
170 & Maria		& 100c	& 120--130	&  \\
\hline
\end{tabular}
\end{table}

%4 & Vesta		& 471	& 425	& cratering \\

%%%%%%%%%%%%%%%%%%%%%%%%%%%%%%%%%%%%%%%%%%%%%%%%%%%%%%%%%%%%%%%%%%%%%%%%

\section{Collisions in the main belt alone}\label{sec:collisions_MB_alone}

Before we proceed to scenarios involving the LHB, we try to explain
the observed families with ages spanning 0--4\,Gyr as a result of collisions only among main-belt bodies.
To this purpose, we used the collisional code called Boulder (Morbidelli et al. 2009)
with the following setup:
the intrinsic probabilities $P_{\rm i} = 3.1\times10^{-18}\,{\rm km}^{-2}\,{\rm yr}^{-1}$,
and the mutual velocities $V_{\rm imp} = 5.28\,{\rm km}/{\rm s}$ for the MB vs MB collisions
(both were taken from the work of Dahlgren 1998).
The assumption of a single $V_{\rm imp}$ value is a simplification,
but about 90\,\% collisions have mutual velocities between $2\hbox{ and }8\,{\rm km}/{\rm s}$ (Dahlgren 1998),
which assures a similar collisional regime.

The scaling law is described by the polynomial relation ($r$~denotes radius in cm)
\begin{equation}
Q^\star_D(r) = {1\over q_{\rm fact}} \left(Q_0 r^a + B \rho r^b\right)\label{Q_star_D}
\end{equation}
with the parameters corresponding to basaltic material at 5\,km/s (Benz \& Asphaug 1999):
\medskip
\begin{center}
\begin{tabular}{cccccc}
\hline
\hline
$\rho$ & $Q_0$ & $a$ & $B$ & $b$ & $q_{\rm fact}$ \\
$({\rm g}/{\rm cm}^3)\!\!\!$ & $\!\!({\rm erg}/{\rm g})\!\!$ & & $\!\!\!({\rm erg}/{\rm g})\!\!\!$ & & \\
\hline
\vrule height 10pt width 0pt
3.0 & $7\times10^7$ & $-0.45$ & 2.1 & 1.19 & 1.0 \\
\hline
\end{tabular}
\end{center}
\medskip
Even though not all asteroids are basaltic, we use the scaling law above
as a mean one for the main-belt population. Below, we discuss also
the case of significantly lower strengths (i.e. higher $q_{\rm fact}$ values).

We selected the time span of the simulation 4\,Gyr (not 4.5\,Gyr)
since we are interested in this last evolutionary phase of the main belt,
when its population and collisional activity is nearly same as today
(Bottke et al. 2005).
The outcome of a single simulation also depends on the ``seed'' value
of the random-number generator that is used in the Boulder code to
decide whether a collision with a fractional probability actually
occurs or not in a given time step.
We thus have to run multiple simulations (usually 100) to obtain information on this
stochasticity of the collisional evolution process.

\begin{figure*}
\centering
\begin{tabular}{@{}c@{}c@{}c@{}}
MB alone & MB vs comets & MB vs disrupting comets \\
\includegraphics[width=6.0cm]{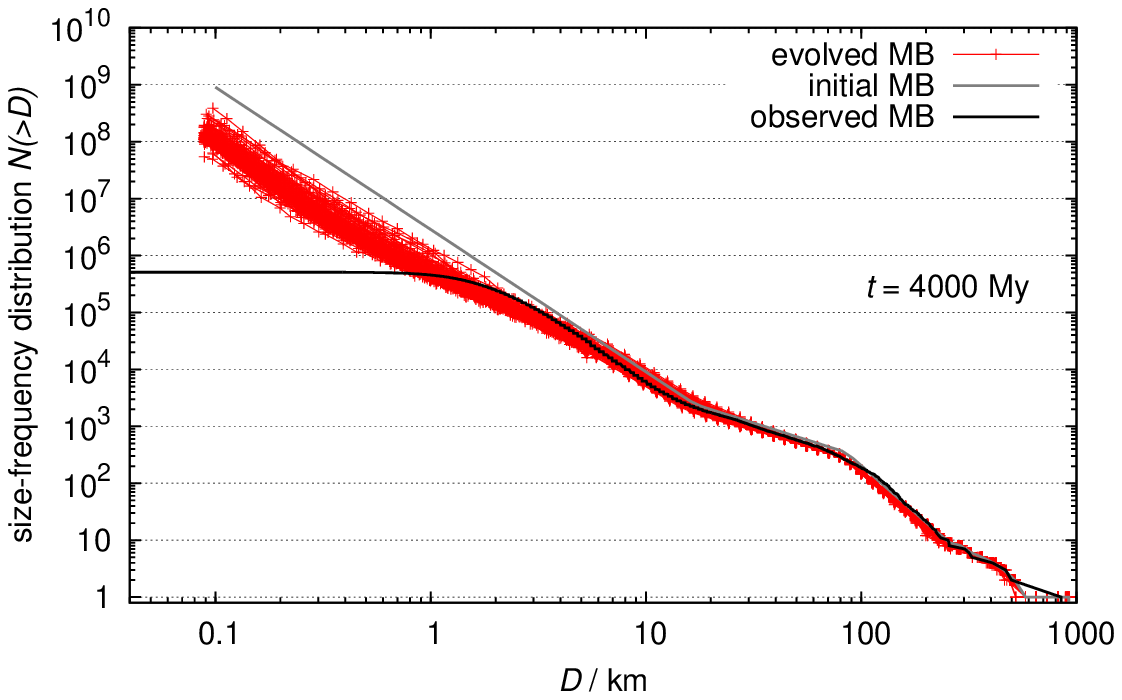} &
\includegraphics[width=6.0cm]{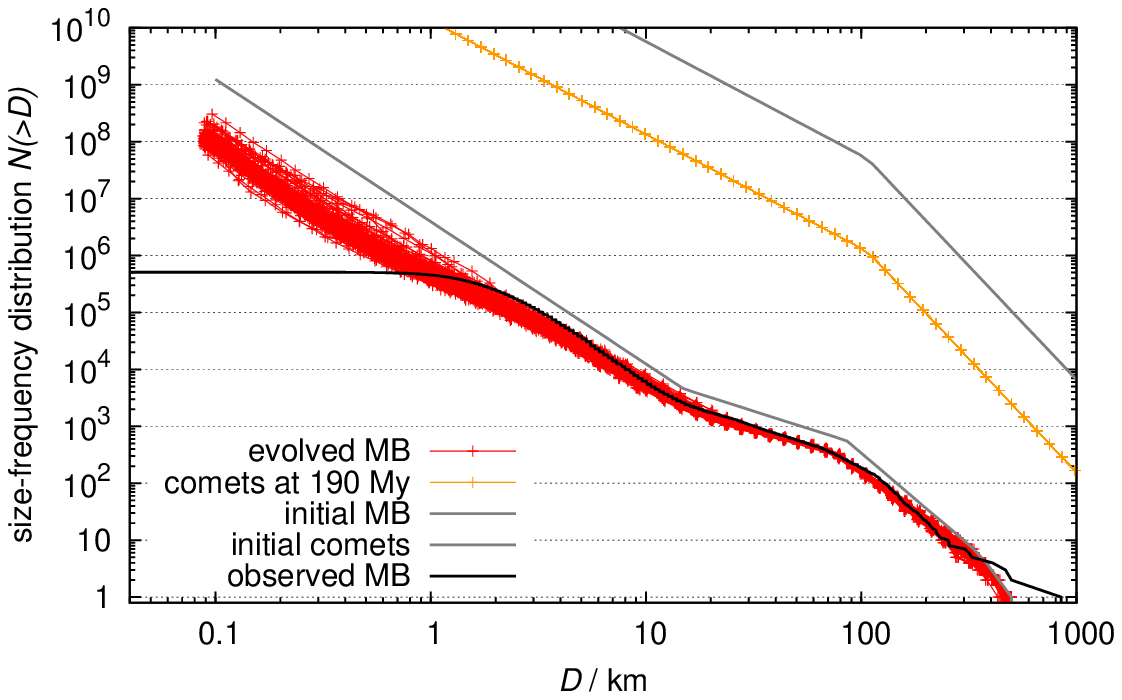} &
\includegraphics[width=6.0cm]{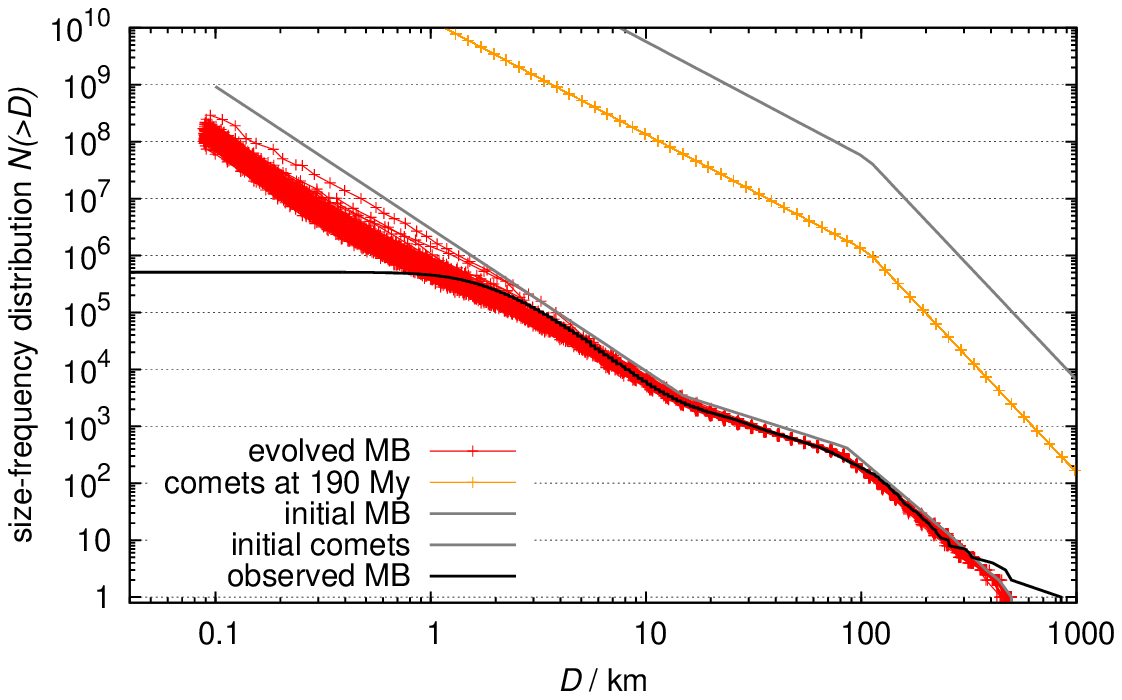} \\
\includegraphics[width=6.0cm]{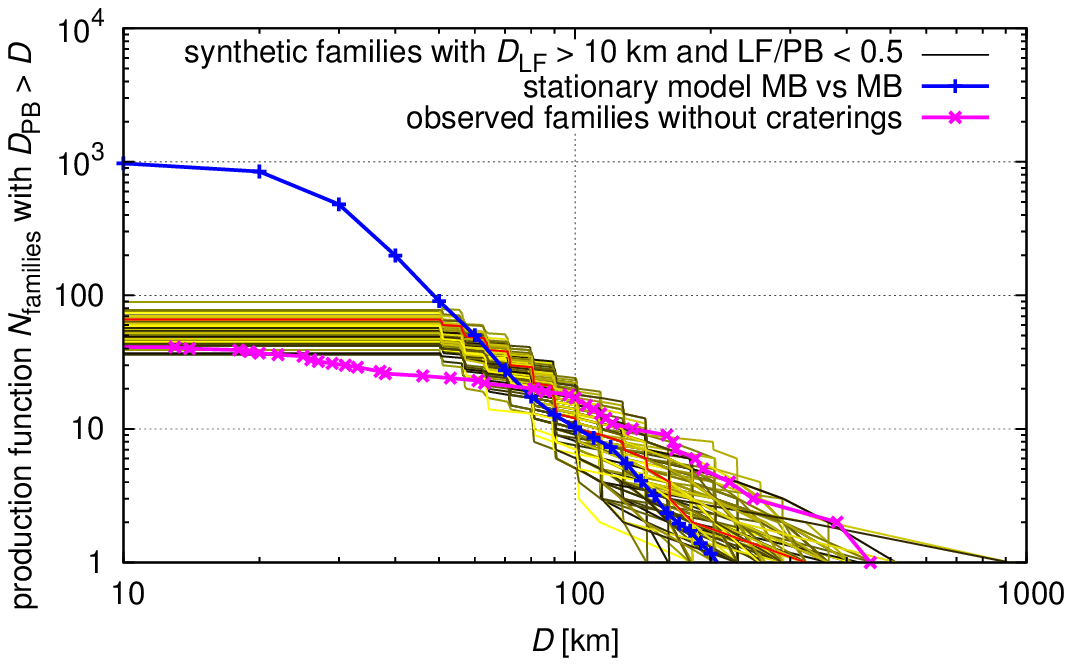} &
\includegraphics[width=6.0cm]{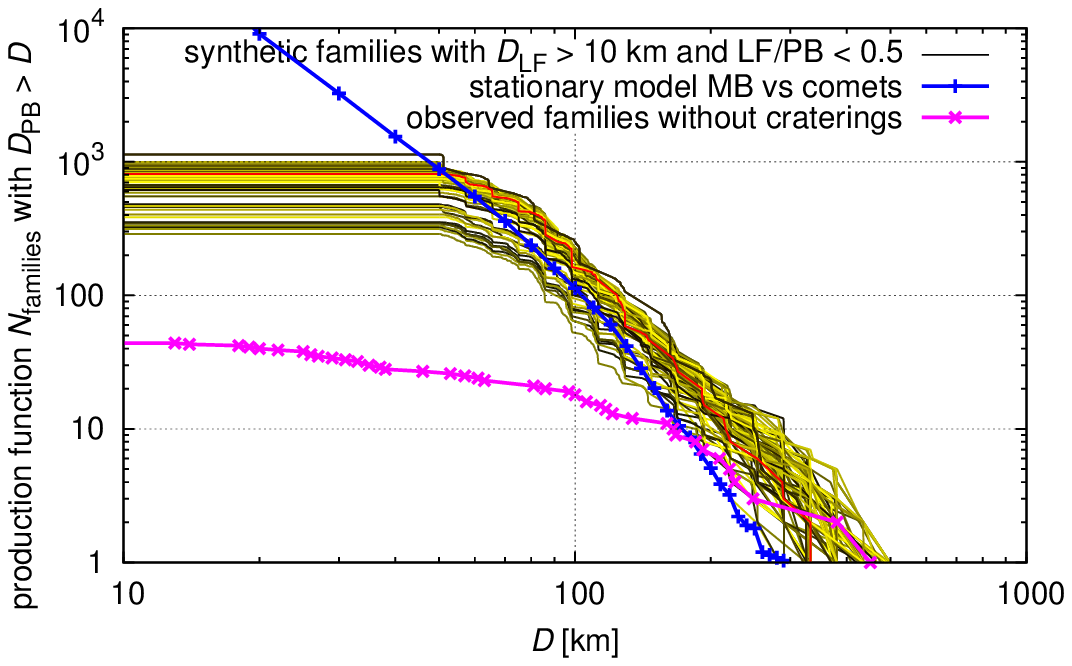} &
\includegraphics[width=6.0cm]{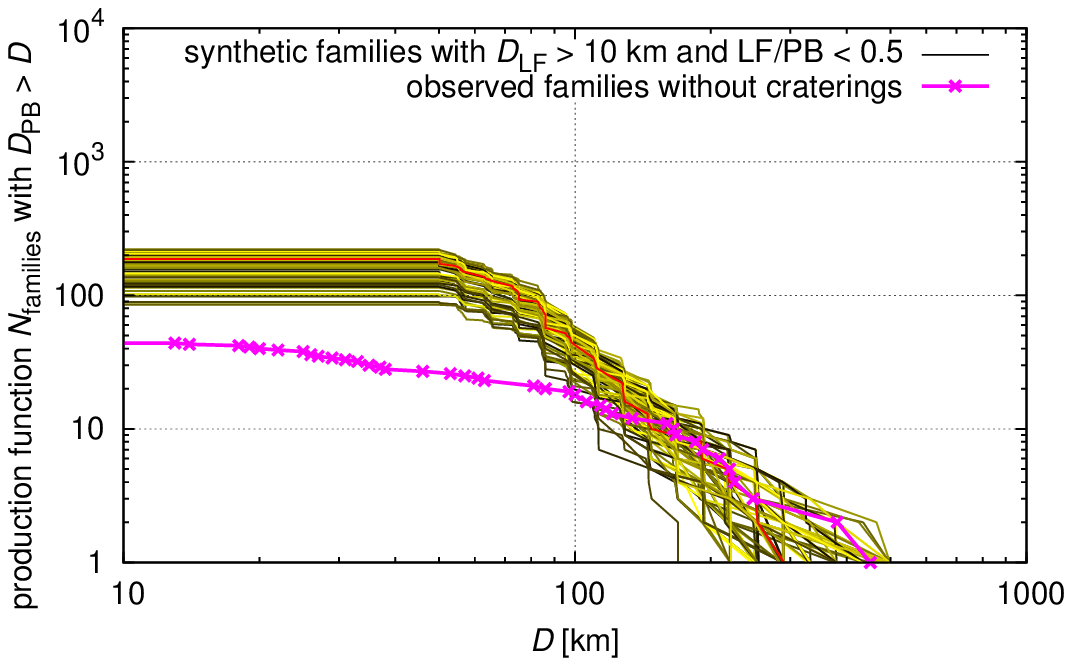} \\
\includegraphics[width=6.0cm]{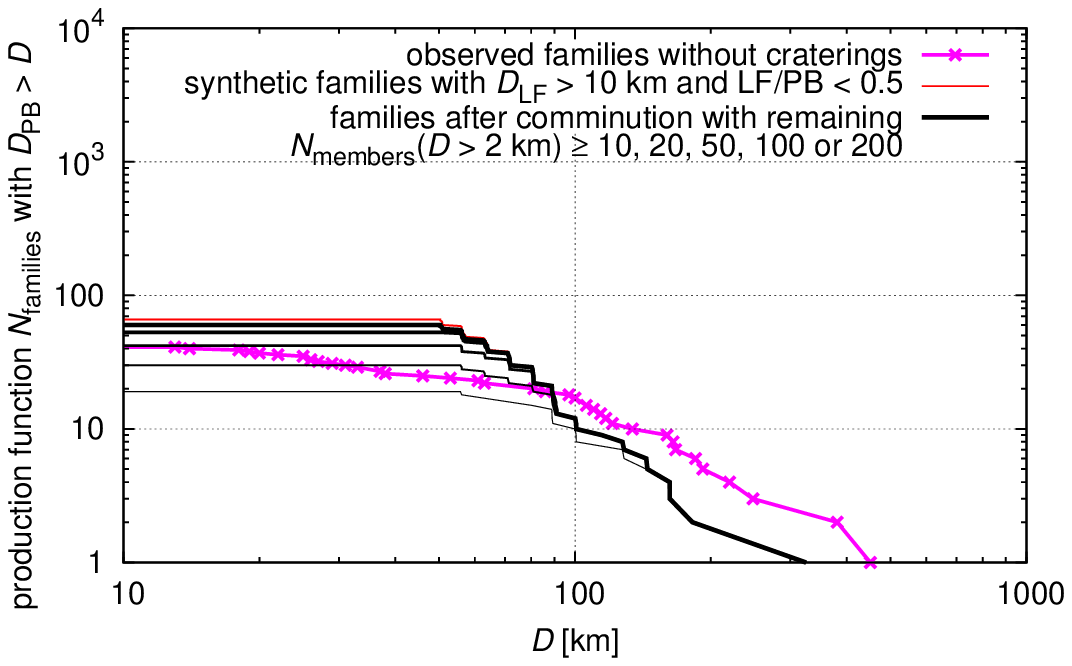} &
\includegraphics[width=6.0cm]{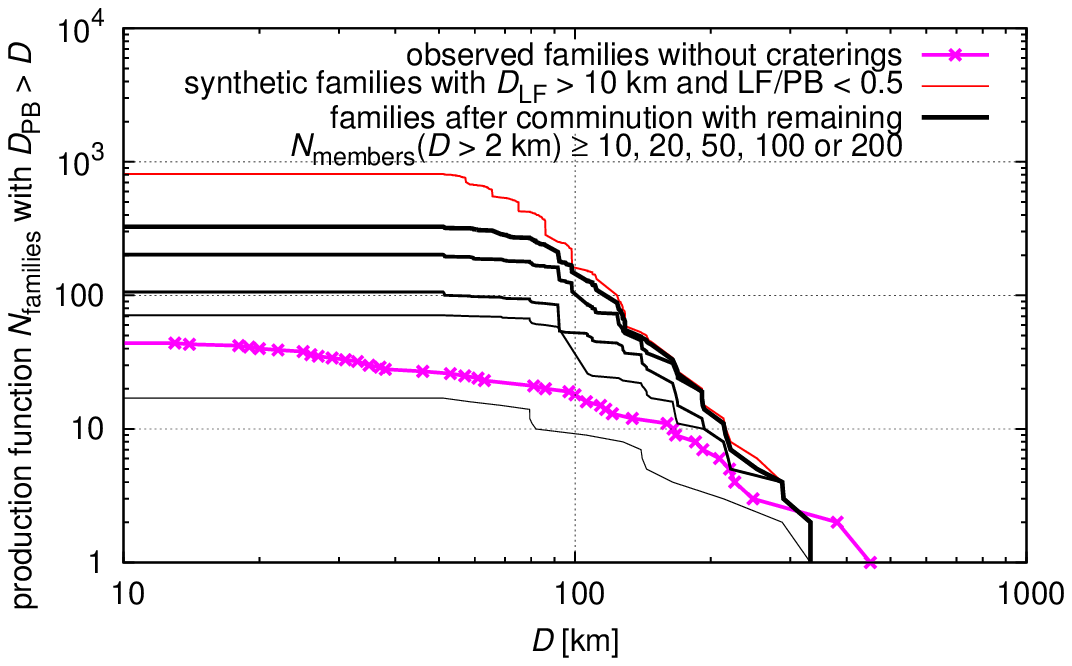} &
\includegraphics[width=6.0cm]{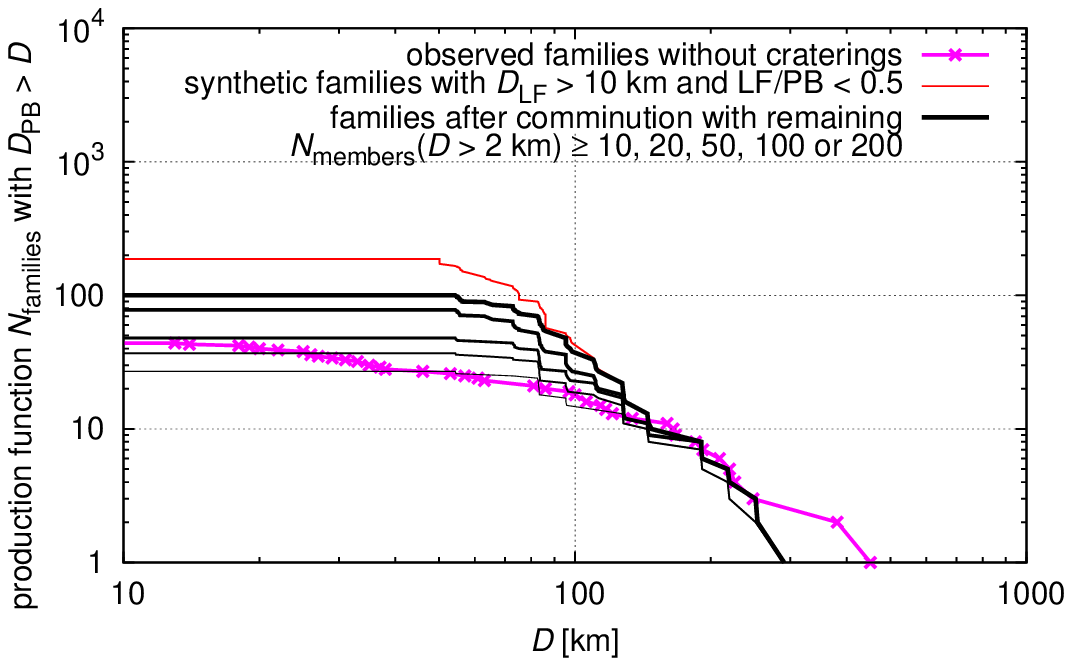} \\
\includegraphics[width=6.0cm]{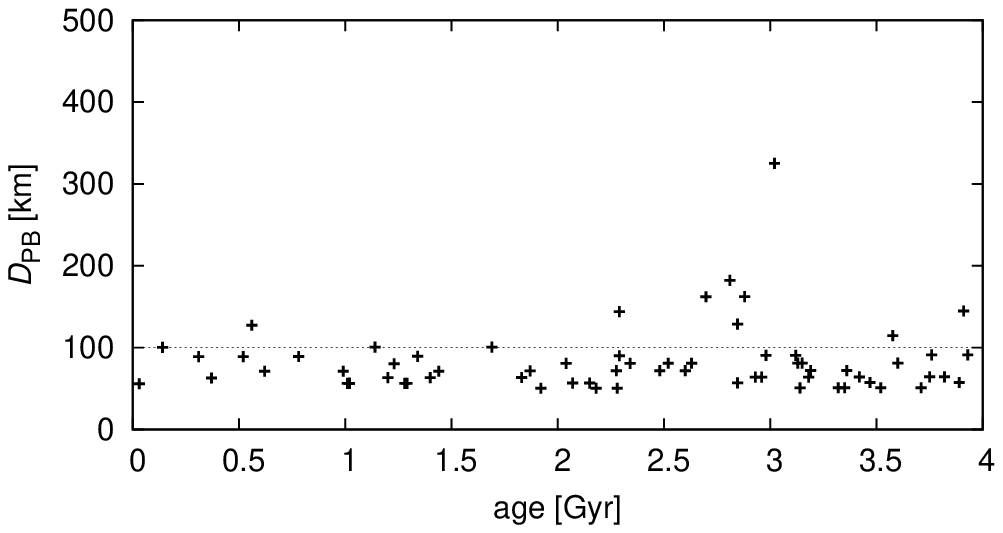} &
\includegraphics[width=6.0cm]{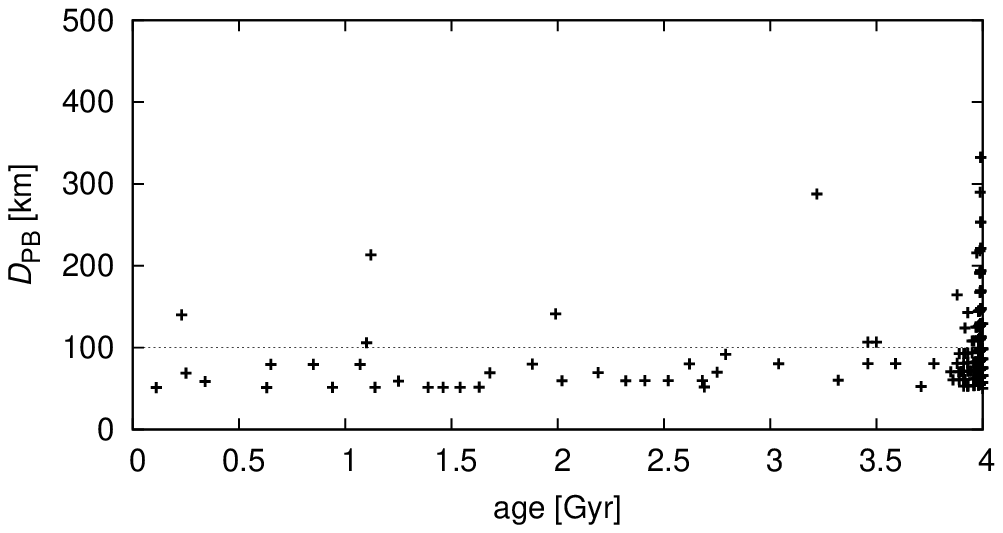} &
\includegraphics[width=6.0cm]{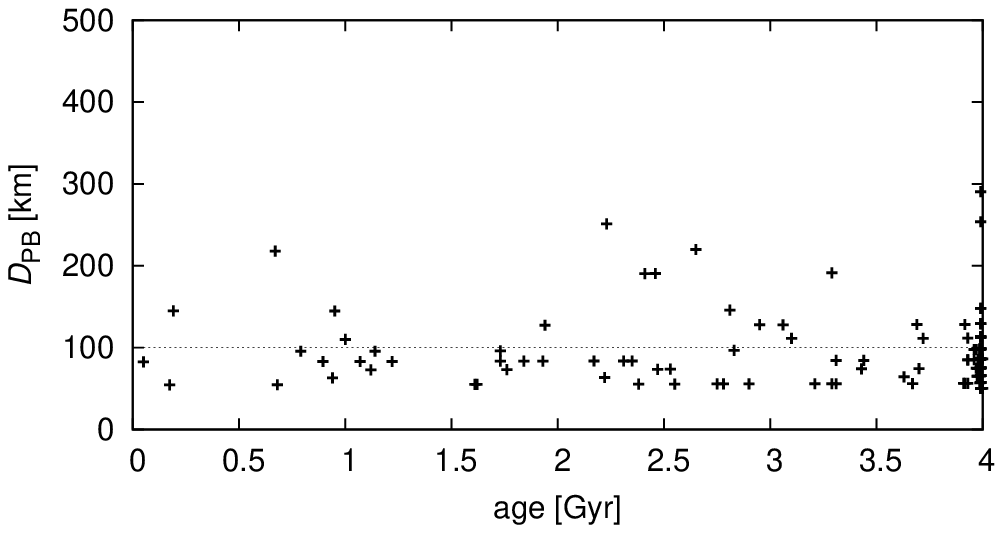} \\
\end{tabular}
\caption{Results of three different collisional models:
main-belt alone which is discussed in Section~\ref{sec:collisions_MB_alone} (left column),
main-belt and comets from Section~\ref{sec:collisions_MB_comets} (middle column),
main-belt and disrupting comets from Section~\ref{sec:comets_lifetime} (right column).
1st row: the initial and evolved size-frequency distributions of the main belt populations for 100 Boulder simulations;
2nd row: the resulting family production functions (in order to distinguish 100 lines we plot them using different colours ranging from black to yellow)
         and their comparison to the observations;
3rd row: the production function affected by comminution for a selected simulation; and
4th row: the distribution of synthetic families with $D_{\rm PB} \ge 50\,{\rm km}$
         in the (age, $D_{\rm PB}$) plot for a selected simulation, without comminution.
The positions of synthetic families in the 4th-row figures may differ significantly
for a different Boulder simulation due to stochasticity and low-number statistics.
Moreover, in the middle and right columns, many families were created during the LHB,
so there are many overlapping crosses close to $4\,{\rm Gyr}$.}
\label{boulder_MB_REAL_SFD_FAMS4_sfd_4000}
\label{boulder_MB_comets_FAMS4_sfd_4000}
\label{boulder_MB_QDISRUPT_FAMS4_sfd_4000}
\end{figure*}

The initial SFD of the main belt population conditions was
approximated by a three-segment power law (see also thin grey line
in Figure~\ref{boulder_MB_REAL_SFD_FAMS4_sfd_4000}, 1st row)
with differential slopes
$q_a = -4.3$ (for $D > D_1$),
$q_b = -2.2$,
$q_c = -3.5$ (for $D < D_2$)
where the size ranges were delimited by $D_1 = 80\,{\rm km}$ and $D_2 = 16\,{\rm km}$.
We also added a few big bodies to reflect the observed shape of the SFD at large sizes ($D > 400\,{\rm km}$).
The normalisation was $N_{\rm norm}(D > D_1) = 350$ bodies in this case.

We used the observed SFD of the main belt as the first constraint
for our collisional model.
We verified that the outcome our model after 4\,Gyr is {\em not\/} sensitive
to the value of~$q_c$. Namely, a change of $q_c$ by as much as $\pm 1$
does not affect the final SFD in any significant way. On the other hand,
the values of the remaining parameters ($q_a$, $q_b$, $D_1$, $D_2$, $N_{\rm norm}$)
are enforced by the observed SFD. To obtain a reasonable fit,
they cannot differ much (by more than 5--10\,\%) from the values presented above.

We do {\em not\/} use only a single number to describe the number
of observed families (e.g. $N = 20$ for $D_{\rm PB} \ge 100\,{\rm km}$),
but we discuss a complete production function instead.
The results in terms of the production function are shown in Figure~\ref{boulder_MB_REAL_SFD_FAMS4_sfd_4000} (left column, 2nd row).
On average, the synthetic production function is steeper and {\em below\/} the observed one,
even though there is approximately a 5\,\% chance that a single realization
of the computer model will resemble the observations quite well.
This also holds for the distribution of $D_{\rm PB} = 200$--400\,km families
in the course of time (age).

In this case, the synthetic production function of $D_{\rm PB} \gtrsim 100\,{\rm km}$ families
is {\em not\/} significantly affected by comminution. According to Bottke et al. (2005),
most of $D > 10\,{\rm km}$ fragments survive intact and a $D_{\rm PB} \gtrsim 100\,{\rm km}$ family
should be recognisable today. This is also confirmed by calculations with Boulder
(see Figure~\ref{boulder_MB_REAL_SFD_FAMS4_sfd_4000}, left column, 3rd row).

To improve the match between the synthetic and the observed production function,
we can do the following:
 i)~modify the scaling law, or
ii)~account for a dynamical decay of the MB population.
Using a substantially lower strength ($q_{\rm fact} = 5$ in Eq.~(\ref{Q_star_D}),
which is not likely, though)
one can obtain a synthetic production function which is {\em on average\/}
consistent with the observations in the $D_{\rm PB} = 200$--400\,km range.

Regarding the dynamical decay, Minton \& Malhotra (2010) suggest that
initially the MB was three times more populous than today while the 
decay timescale was very short: after 100 Myr of evolution the number
of bodies is almost at the current level. In this brief period of time,
about 50\,\% more families will be created, but all of them will be old, of course.
For the remaining $\sim 3.9\,{\rm Gyr}$, the above model (without any dynamical decay) is valid. 

To conclude, it is possible -- though not very likely -- that the observed families
were produced by the collisional activity in the main belt alone.
A dynamical decay of the MB population would create more families that are old,
but technically speaking, this cannot be distinguished from the LHB scenario,
which is discussed next.

% ***SAY WHAT ARE THE BLUE, YELLOW AND ORANGE LINES

%%%%%%%%%%%%%%%%%%%%%%%%%%%%%%%%%%%%%%%%%%%%%%%%%%%%%%%%%%%%%%%%%%%%%%%%

\section{Collisions between a ``classical'' cometary disk and the main belt}\label{sec:collisions_MB_comets}

In this section, we construct a collisional model and estimate
an expected number of families created during the LHB due to collisions
between cometary-disk bodies and main-belt asteroids. We start with a simple stationary model
and we confirm the results using a more sophisticated Boulder code (Morbidelli et al. 2009).

\begin{figure}
\centering
\includegraphics[width=8cm]{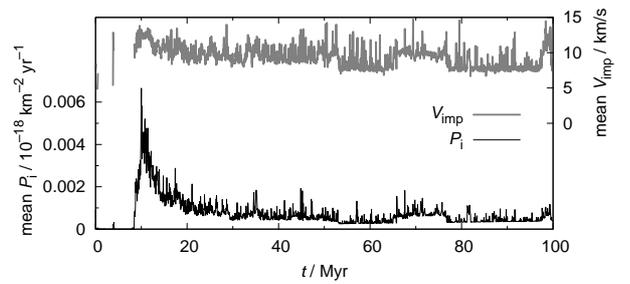}
\caption{The temporal evolution of the intrinsic collisional probability $P_{\rm i}$ (bottom)
and mean collisional velocity $V_{\rm imp}$ (top) computed for collisions
between cometary-disk bodies and the main-belt asteroids. The time $t = 0$ is arbitrary here;
the sudden increase in $P_i$ values corresponds to the beginning of the LHB.}
\label{probat_impvel}
\end{figure}

Using the data from Vokrouhlick\'y et al. (2008) for a ``classical'' cometary disk,
we can estimate the intrinsic collisional probability and the collisional velocity
between comets and asteroids. A~typical time-dependent evolution of $P_{\rm i}$ and $V_{\rm imp}$
is shown in Figure~\ref{probat_impvel}. The probabilities increase at first, as the transneptunian
cometary disk starts to decay, reaching up to $6\times 10^{-21}\,{\rm km}^{-2}\,{\rm yr}^{-1}$,
and after $100\,{\rm Myr}$ they decrease to zero. These results do {\em not\/} differ significantly
from run to run.

%%%%%%%%%%%%%%%%%%%%%%%%%%%%%%%%%%%%%%%%%%%%%%%%%%%%%%%%%%%%%%%%%%%%%%%%

\subsection{Simple stationary model}\label{sec:stationary}

In a stationary collisional model,
we choose an SFD for the cometary disk,
we assume a {\em current\/} population of the main belt;
estimate the projectile size needed to disrupt a given target according to (Bottke et al. 2005)
\begin{equation}
d_{\rm disrupt} = \left({2 Q^\star_D / V_{\rm imp}^2}\right)^{1/3} D_{\rm target}\,,\label{eq:d_disrupt}
\end{equation}
where $Q^\star_D$~denotes the specific energy for disruption and
dispersion of the target (Benz \& Asphaug 1999);
and finally calculate the number of events during the LHB as
\begin{equation}
n_{\rm events} = {D_{\rm target}^2\over 4}\, n_{\rm target} \int P_i(t)\, n_{\rm project}(t)\, {\rm d}t \,,
\end{equation}
where $n_{\rm target}$ and $n_{\rm project}$ are the number of targets (i.e. main belt asteroids)
and the number of projectiles (comets), respectively.
The actual number of bodies (27,000) in the dynamical simulation of Vokrouhlick\'y et al. (2008)
changes in the course of time, and it was scaled such that it was initially equal
to the number of projectiles~$N({>}d_{\rm disrupt})$ inferred from the SFD of the disk.
This is clearly a {\em lower limit\/} for the number of families created,
since the main belt was definitely more populous in the past.

The average impact velocity is $V_{\rm imp} \simeq 10\,{\rm km}/{\rm s}$,
so we need the following projectile sizes to disrupt given target sizes:

\begin{center}
\begin{tabular}{ccccc}
\hline
\hline
\vrule height 10pt width 0pt
$D_{\rm target}$ & $N_{\rm targets}$ & $Q^\star_D$ & $d_{\rm disrupt}$ for ${\rho_{\rm target}\over\rho_{\rm project}} = 3 \hbox{ to } 6$ \\
(km) & in the MB & (J/kg) & (km) \\
\hline
\vrule height 10pt width 0pt
100 & $\sim$192 & $1\times10^5$ & 12.6 to 23 \\
200 &  $\sim$23 & $4\times10^5$ & 40.0 to 73 \\
\hline
\end{tabular}
\end{center}

\begin{figure}
\centering
\includegraphics[width=8cm]{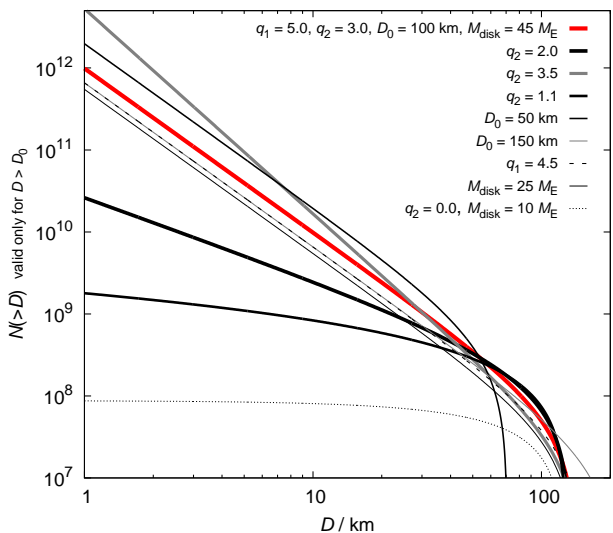}
\caption{Cumulative size-frequency distributions of the cometary disk
tested in this work. All the parameters of our
nominal choice are given in the top label; the other labels
just report the parameters that changed relative to our nominal choice.}
\label{fig:disk_sfd}
\end{figure}

We tried to use various SFDs for the cometary disk (i.e., with various differential slopes~$q_1$ for $D > D_0$ and $q_2$ for $D < D_0$,
the elbow diameter~$D_0$ and total mass~$M_{\rm disk}$), including rather extreme cases (see Figure~\ref{fig:disk_sfd}).
The resulting numbers of LHB families are summarised in Table~\ref{tab:stationary}.
Usually, we obtain several families with $D_{\rm PB} \simeq 200\,{\rm km}$
and about $100$~families with $D_{\rm PB} \simeq 100\,{\rm km}$.
This result is robust with respect to the slope $q_2$,
because even very shallow SFDs should produce a lot of these families.%
\footnote{The extreme case with $q_2 = 0$ is not likely at all,
e.g. because of the continuous SFD of basins on Iapetus and Rhea,
which only exhibits a mild depletion of $D \simeq 100\,{\rm km}$ size craters; see Kirchoff \& Schenk (2010). 
On the other hand, Sheppard \& Trujillo (2010) report an extremely
shallow cumulative SFD of Neptune Trojans that is akin to low~$q_2$.}
The only way to decrease the number of families significantly
is to assume the elbow at a larger diameter~$D_0 \simeq 150\,{\rm km}$.

It is thus no problem to explain the existence of approximately five {\em large\/} families
with $D_{\rm PB} = 200$--400\,km, which are indeed observed, since they can be
readily produced during the LHB.
On the other hand, the high number of $D_{\rm PB} \simeq 100\,{\rm km}$ families clearly contradicts the observations,
since we observe {\em almost no LHB families\/} of this size.

\begin{table*}
\caption{Results of a stationary collisional model between the cometary disk and the main belt.
The parameters characterise the SFD of the disk:
$q_1$, $q_2$~are differential slopes for the diameters larger/smaller than the elbow diameter~$D_0$,
$M_{\rm disk}$~denotes the total mass of the disk,
and $n_{\rm events}$~is the resulting number of families created during the LHB
for a given parent body size~$D_{\rm PB}$.
The ranges of $n_{\rm events}$ are due to variable density ratios $\rho_{\rm target}/\rho_{\rm project} = 1 \hbox{ to } 3/1$.}
\label{tab:stationary}
\begin{tabular}{cccccccl}
\hline
\hline
$q_1$ & $q_2$ & $D_0$ & $M_{\rm disk}$ & $n_{\rm events}$ & & & notes \\
 & & (km) & ($M_\oplus$) & for $D_{\rm PB} \ge 100\,{\rm km}$ & $D_{\rm PB} \ge 200\,{\rm km}$ & Vesta craterings & \\
\hline
5.0 & 3.0 & 100 & 45 & 115--55  & 4.9--2.1  & 2.0  & nominal case \\
5.0 & 2.0 & 100 & 45 &  35--23  & 4.0--2.2  & 1.1  & shallow SFD \\
5.0 & 3.5 & 100 & 45 & 174--70  & 4.3--1.6  & 1.8  & steep SFD \\
5.0 & 1.1 & 100 & 45 &  14--12  & 3.1--2.1  & 1.1  & extremely shallow SFD \\
4.5 & 3.0 & 100 & 45 &  77--37  & 3.3--1.5  & 1.3  & lower $q_1$ \\
5.0 & 3.0 &  50 & 45 & 225--104 & 7.2--1.7  & 3.2  & smaller turn-off \\
5.0 & 3.0 & 100 & 25 &  64--40  & 2.7--1.5  & 1.1  & lower $M_{\rm disk}$ \\
5.0 & 3.0 & 100 & 17 &  34      & 1.2       & 1.9  & $\rho_{\rm comets} = 500\,{\rm kg}/{\rm m}^3$ \\
5.0 & 3.0 & 150 & 45 &  77--23  & 3.4--0.95 & 0.74 & larger turn-off \\
5.0 & 0.0 & 100 & 10 & 1.5--1.4 & 0.5--0.4  & 0.16 & worst case (zero $q_2$ and low $M_{\rm disk}$) \\
\hline
\end{tabular}
\end{table*}

%%%%%%%%%%%%%%%%%%%%%%%%%%%%%%%%%%%%%%%%%%%%%%%%%%%%%%%%%%%%%%%%%%%%%%%%

\subsection{Constraints from (4) Vesta}\label{sec:vesta}

The asteroid (4) Vesta presents a significant constraint for collisional models,
being a differentiated body with a preserved basaltic crust (Keil 2002)
and a 500\,km large basin on its surface (a feature indicated by the photometric analysis of Cellino et al. 1987),
which is significantly younger than 4\,Gyr (Marchi et al. 2012).
It is highly unlikely that Vesta experienced a catastrophic disruption in the past,
and even large cratering events were limited.
We thus have to check the number of collisions between one $D = 530$\,km target and $D \simeq 35$\,km projectiles,
which are capable of producing the basin and the Vesta family (Thomas et al. 1997). According to Table~\ref{tab:stationary},
the predicted number of such events does not exceed $\sim2$, so given the stochasticity of the results
there is a significant chance that Vesta indeed experienced zero
such impacts during the LHB.

%%%%%%%%%%%%%%%%%%%%%%%%%%%%%%%%%%%%%%%%%%%%%%%%%%%%%%%%%%%%%%%%%%%%%%%%

\subsection{Simulations with the Boulder code}\label{sec:boulder}

To confirm results of the simple stationary model, we also performed
simulations with the Boulder code. We modified the code to include a time-dependent
collisional probability~$P_{\rm i}(t)$ and impact velocity~$V_{\rm imp}(t)$
of the cometary-disk population.

We started a simulation with a setup for the cometary disk resembling the nominal case from Table~\ref{tab:stationary}.
The scaling law is described by Eq.~(\ref{Q_star_D}),
with the following parameters (the first set corresponds to basaltic material at 5\,km/s,
the second one to water ice, Benz \& Asphaug 1999):

\begin{center}
\begin{tabular}{ccccccc}
\hline
\hline
		& $\rho$ & $Q_0$ & $a$ & $B$ & $b$ & $q_{\rm fact}$ \\
		& $\!\!\!({\rm g}/{\rm cm}^3)\!\!\!$ & $\!\!({\rm erg}/{\rm g})\!\!$ & & $\!\!\!({\rm erg}/{\rm g})\!\!\!$ & & \\
\hline
\vrule height 10pt width 0pt
asteroids	& 3.0 & $7\times10^7$   & $-0.45$ & 2.1 & 1.19 & 1.0 \\
comets		& 1.0 & $1.6\times10^7$ & $-0.39$ & 1.2 & 1.26 & 3.0 \\
\hline
\end{tabular}
\end{center}

The intrinsic probabilities $P_{\rm i} = 3.1\times10^{-18}\,{\rm km}^{-2}\,{\rm yr}^{-1}$
and velocities $V_{\rm imp} = 5.28\,{\rm km}/{\rm s}$ for the MB vs MB collisions
were again taken from the work of Dahlgren (1998). We do not account for comet--comet
collisions since their evolution is dominated by the dynamical decay.
The initial SFD of the main belt was similar to the one in Section~\ref{sec:collisions_MB_alone},
$q_a = -4.2$,
$q_b = -2.2$,
$q_c = -3.5$,
$D_1 = 80\,{\rm km}$,
$D_2 = 14\,{\rm km}$,
and only the normalisation was increased up to $N_{\rm norm}(D > D_1) = 560$ in this case.

The resulting size-frequency distributions of 100 independent simulations with different
random seeds are shown in Figure~\ref{boulder_MB_comets_FAMS4_sfd_4000} (middle column).
The number of LHB families (approximately~10 with $D_{\rm PB} \simeq 200\,{\rm km}$ and 200 with $D_{\rm PB} \simeq 100\,{\rm km}$)
is even {\em larger\/} compared to the stationary model, as expected,
because we had to start with a larger main belt to get a good fit
of the currently observed MB after 4\,Gyr of collisional evolution.

To conclude, the stationary model and the Boulder code give results
that are compatible with each other, but that clearly contradict the
observed production function of families. In particular, they predict
far too many families with $D=100$\,km parent bodies. At first sight,
this may be interpreted as proof that there was no cometary LHB on
the asteroids. Before jumping to this conclusion, however, one has to
investigate whether there are biases against identifying of
$D_{\rm PB} = 100\,{\rm km}$ families. In Sections~\ref{sec:overlap}--\ref{sec:comminution}
we discuss several mechanisms that all contribute, at some level, to reducing the number
of observable $D_{\rm PB} = 100$\,km families over time.
They are addressed in order of relevance, from the least to the most effective.

%%%%%%%%%%%%%%%%%%%%%%%%%%%%%%%%%%%%%%%%%%%%%%%%%%%%%%%%%%%%%%%%%%%%%%%%

\section{Families overlap}\label{sec:overlap}

Because the number of expected $D_{\rm PB} \ge 100\,{\rm km}$ LHB families is very high (of the order of $100$)
we now want to verify if these families can {\em overlap\/} in such a way that they
cannot be distinguished from each other and from the background.
We thus took 192 main-belt bodies with $D \ge 100\,{\rm km}$
and selected randomly 100~of them that will break apart. For each one we created
an artificial family with $10^2$~members,
assume a size-dependent ejection velocity $V \propto 1/D$ (with $V = 50\,{\rm m}/{\rm s}$ for $D = 5\,{\rm km}$)
and the size distribution resembling that of the Koronis family.
The choice of the true anomaly and the argument of perihelion at the
instant of the break-up event was random.
We then calculated proper elements~$(a_{\rm p}, e_{\rm p}, \sin I_{\rm p})$ for all bodies.
This type of analysis is in some respects similar to the work of Bendjoya et al. (1993).

\begin{figure}
\centering
\includegraphics[width=8cm]{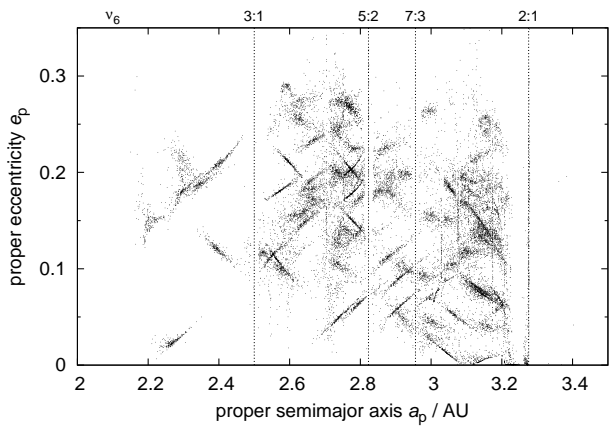}
\caption{The proper semimajor axis~$a_{\rm p}$ vs the proper eccentricity~$e_{\rm p}$
for 100~synthetic families created in the main belt. It is the {\em initial\/} state,
shortly after disruption events. We assume the size-frequency distribution of bodies
in each synthetic family similar to that of the Koronis family (down to $D \simeq 2\,{\rm km}$).
Break-ups with the true anomaly $f \simeq 0\hbox{ to }30^\circ$ and $150^\circ\hbox{ to }180^\circ$
are more easily visible on this plot, even though the choice of both~$f$
and the argument of perihelion $\varpi$ was random for all families.}
\label{fams-1_SMALLD_ae}
\end{figure}

According to the resulting Figure~\ref{fams-1_SMALLD_ae} the answer to the question
is simple: the families do {\em not\/} overlap sufficiently, and they cannot be hidden that way.
Moreover, if we take only bigger bodies ($D > 10\,{\rm km}$), these would be clustered
even more tightly. The same is true for proper inclinations, which are usually
more clustered than eccentricities, so families could be more easily
recognised.

%%%%%%%%%%%%%%%%%%%%%%%%%%%%%%%%%%%%%%%%%%%%%%%%%%%%%%%%%%%%%%%%%%%%%%%%

\section{Dispersion of families by the Yarkovsky drift}\label{sec:yarko_dispersal}

In this section, we model long-term evolution of synthetic families driven
by the Yarkovsky effect and chaotic diffusion. For {\em one\/} synthetic family
located in the outer belt, we have performed a full N-body integration with the SWIFT package (Levison \& Duncan 1994),
which includes also an implementation of the Yarkovsky/YORP effect (Bro\v z 2006)
and second-order integrator by Laskar \& Robutel (2001). We included 4~giant planets in this simulation.
To speed-up the integration, we used ten times smaller sizes of the test particles
and thus a ten times shorter time span (400\,Myr instead of 4\,Gyr). The selected time step is $\Delta t = 91\,{\rm d}$.
We computed proper elements, namely their differences $\Delta a_{\rm p}, \Delta e_{\rm p}, \Delta \sin I_{\rm p}$
between the initial and final positions.

Then we used a simple {\em Monte-Carlo\/} approach for the whole set of 100 synthetic families
--- we assigned a suitable drift~$\Delta a_{\rm p}(D)$ in semimajor axis,
and also drifts in eccentricity~$\Delta e_{\rm p}$ and inclination~$\Delta \sin I_{\rm p}$
to each member of 100 families, respecting asteroid sizes, of course. 
This way we account for the Yarkovsky semimajor axis drift and also for interactions
with mean-motion and secular resonances.
This Monte-Carlo method tends to smear all structures,
so we can regard our results as the {\em upper limits\/} for dispersion of families.

\begin{figure}
\centering
\includegraphics[width=8cm]{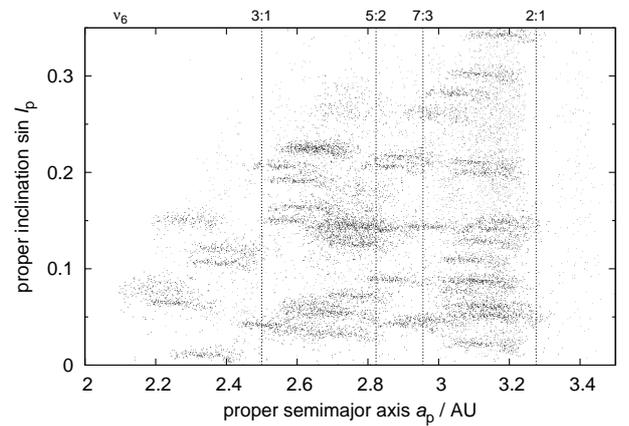}
\caption{The proper semimajor axis $a_{\rm p}$ vs the proper inclination $\sin I_{\rm p}$
for 100~synthetic asteroid families (black dots), {\em evolved\/} over 4\,Gyr using a Monte-Carlo model.
The assumed SFDs correspond to the Koronis family,
but we show only $D > 10\,{\rm km}$ bodies here.
We also include $D > 10\,{\rm km}$ background asteroids (grey dots) for comparison.}
\label{fams-1_ai_YE_MonteCarlo_017_BACKGROUND}
\end{figure}

While the eccentricities of small asteroids (down to $D \simeq 2\,{\rm km}$) seem to be dispersed enough to hide the families,
there are still some persistent structures in inclinations, which would be observable today.
Moreover, large asteroids ($D \ge 10\,{\rm km}$) seem to be clustered even after $4\,{\rm Gyr}$,
so that more than 50\,\% of families can be easily recognised against the background
(see Figure~\ref{fams-1_ai_YE_MonteCarlo_017_BACKGROUND}).
We thus can conclude that it is {\em not\/} possible to disperse the families by the Yarkovsky effect alone.

%%%%%%%%%%%%%%%%%%%%%%%%%%%%%%%%%%%%%%%%%%%%%%%%%%%%%%%%%%%%%%%%%%%%%%%%

\section{Reduced physical lifetime of comets in the MB crossing zone}\label{sec:comets_lifetime}

To illustrate the effects that the physical disruption of comets
(due to volatile pressure build-up, amorphous/crystalline phase transitions, spin-up by jets, etc.)
can have on the collisional evolution of the asteroid belt, we adopted here a simplistic assumption.
We considered that no comet disrupt beyond 1.5\,AU, whereas all comets
disrupt the first time that they penetrate inside 1.5\,AU. Both
conditions are clearly not true in reality: some comets are observed
to blow up beyond 1.5\,AU, and others are seen to survive on an Earth-crossing
orbit. Thus we adopted our disruption law just as an example 
of a drastic reduction of the number of comets with small perihelion
distance, as required to explain the absence of evidence for a
cometary bombardment on the Moon. 

We then removed all those objects from output of comet evolution during the LHB
that had a passage within 1.5\,AU from the Sun,
from the time of their first passage below this threshold.
We then recomputed the mean intrinsic collision probability of a comet
with the asteroid belt. The result is a factor $\sim$3 smaller than when no
physical disruption of comets is taken into account as in Fig.~\ref{probat_impvel}.
The mean impact velocity with asteroids also decreases, from 12\,km/s to
8\,km/s.  

The resulting number of asteroid disruption events is thus decreased
by a factor $\sim$4.5, which can be also seen in the production
function shown in Figure~\ref{boulder_MB_QDISRUPT_FAMS4_sfd_4000} (right column).
The production of families with $D_{\rm PB} = 200$--400\,km is consistent
with observations, while the number of $D_{\rm PB} \simeq 100$\,km
families is reduced to 30--70, but is still too high, by a factor 2--3.
More importantly, the slope of the production function remains steeper
than that of the observed function. Thus, our conclusion is that
physical disruptions of comets alone {\em cannot\/} explain the
observation, but may be an important factor to keep in mind for
reconciling the model with the data.

%%%%%%%%%%%%%%%%%%%%%%%%%%%%%%%%%%%%%%%%%%%%%%%%%%%%%%%%%%%%%%%%%%%%%%%%

\begin{figure*}
\centering
\includegraphics[width=7cm]{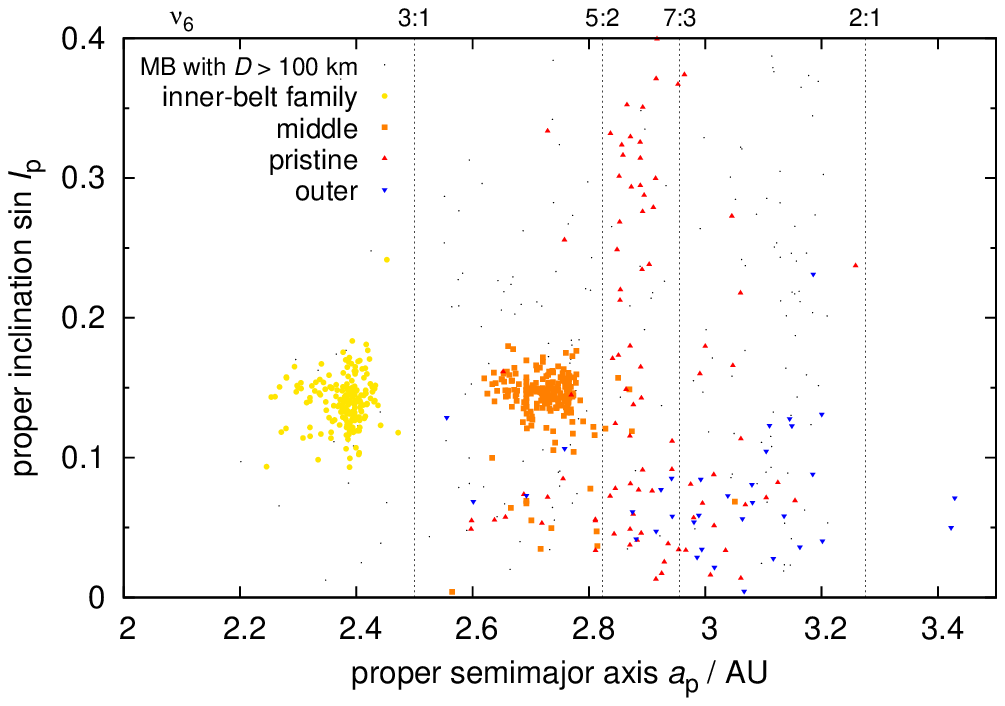}
\kern.3cm
\includegraphics[width=7cm]{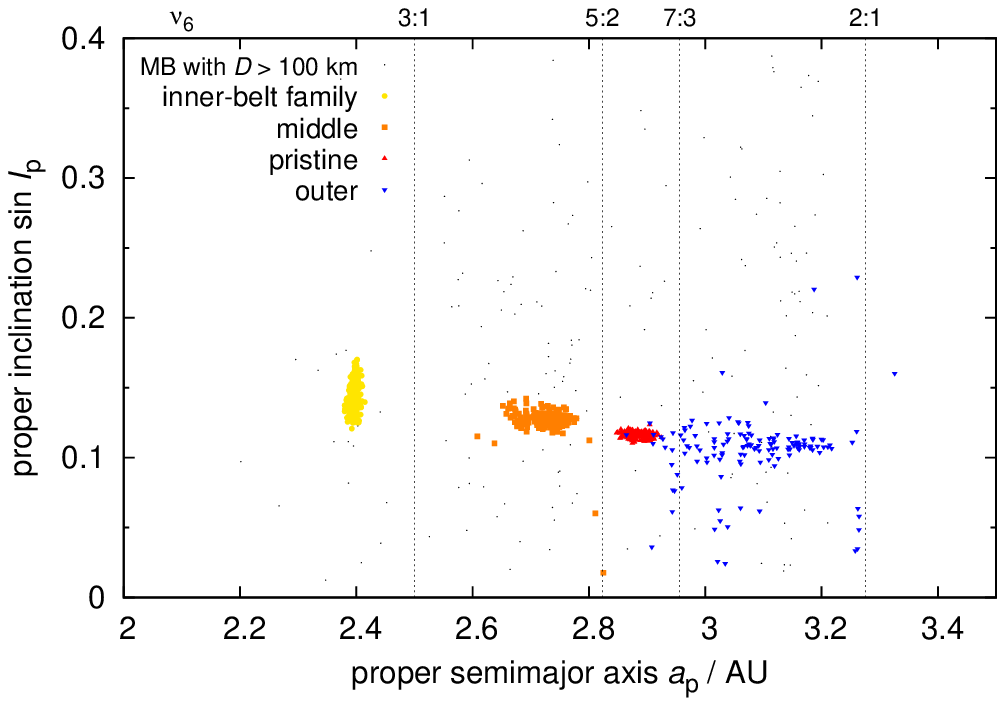}
\caption{The proper semimajor axis vs the proper inclination for four synthetic
families (distinguished by symbols) as perturbed by giant-planet migration.
Left panel: the case when families were evolved over the ``jump'' due to the encounter between Jupiter and Neptune.
Right panel: the families created just after the jump and perturbed only by later phases of migration.}
\label{jumping_jupiter_fams_1_T0_0My_ai}
\end{figure*}

\section{Perturbation of families by migrating planets (a~jumping-Jupiter scenario)}\label{sec:jumping_jupiter}

In principle, families created during the LHB may be perturbed by still-migrating planets.
It is an open question what the exact orbital evolution of planets was at that time.
Nevertheless, a plausible scenario called a ``jumping Jupiter'' was presented by Morbidelli et al. (2010).
It explains major features of the main belt (namely the paucity of high-inclination asteroids
above the $\nu_6$ secular resonance), and is consistent with amplitudes of the secular frequencies of both giant and terrestrial planets
and also with other features of the solar system. In this work, we thus investigated this particular
migration scenario.

We used the data from Morbidelli et al. (2010) for the orbital evolution of giant planets.
We then employed a modified SWIFT integrator, which read orbital elements for planets from an input file
and calculated only the evolution of test particles.
Four synthetic families located in the inner/middle/outer belt were integrated.
We started the evolution of planets at various times, ranging from $t_0$
to ($t_0 + 4\,{\rm Myr}$) and stopped the integration at ($t_0 + 4\,{\rm My}$),
in order to test the perturbation on families created in different phases of migration.
Finally, we calculated proper elements of asteroids when the planets do not migrate anymore.
(We also had to move planets smoothly to their exact current orbital positions.)

The results are shown in Figure~\ref{jumping_jupiter_fams_1_T0_0My_ai}.
While the proper eccentricities seem to be sufficiently perturbed
and families are dispersed even when created at late phases of migration,
the proper inclinations are not very dispersed, except for families in
the outer asteroid belt that formed at the very beginning of the giant
planet instability (which may be unlikely, as there must be a delay
between the onset of planet instability and the beginning of the
cometary flux through the asteroid belt). In most cases, the LHB families could
still be identified as clumps in semi-major axis vs inclination space. 
We do not see any of such ($a_{\rm p}, \sin I_{\rm p}$)-clumps, dispersed in eccentricity,
in the asteroid belt.%
\footnote{High-inclination families would be dispersed much
more owing to the Kozai mechanism, because eccentricities that are sufficiently perturbed
exhibit oscillations coupled with inclinations.}

The conclusion is clear: it is {\em not\/} possible to destroy low-$e$ and low-$I$ families
by perturbations arising from giant-planet migration, at least in the case
of the ``jumping-Jupiter'' scenario.%
\footnote{The currently non-existent families around (107)~Camilla and (121)~Hermione ---
inferred from the existence of their satellites --- cannot be destroyed
in the jumping-Jupiter scenario, unless the families were actually {\em pre\/}-LHB
and had experienced the jump.}

%%%%%%%%%%%%%%%%%%%%%%%%%%%%%%%%%%%%%%%%%%%%%%%%%%%%%%%%%%%%%%%%%%%%%%%%

\section{Collisional comminution of asteroid families}\label{sec:comminution}

We have already mentioned that the comminution is {\em not\/} sufficient to destroy
a $D_{\rm PB} = 100\,{\rm km}$ family {\em in the current environment\/}
of the main belt (Bottke et al. 2005).

\begin{figure*}
\centering
\includegraphics[width=6cm]{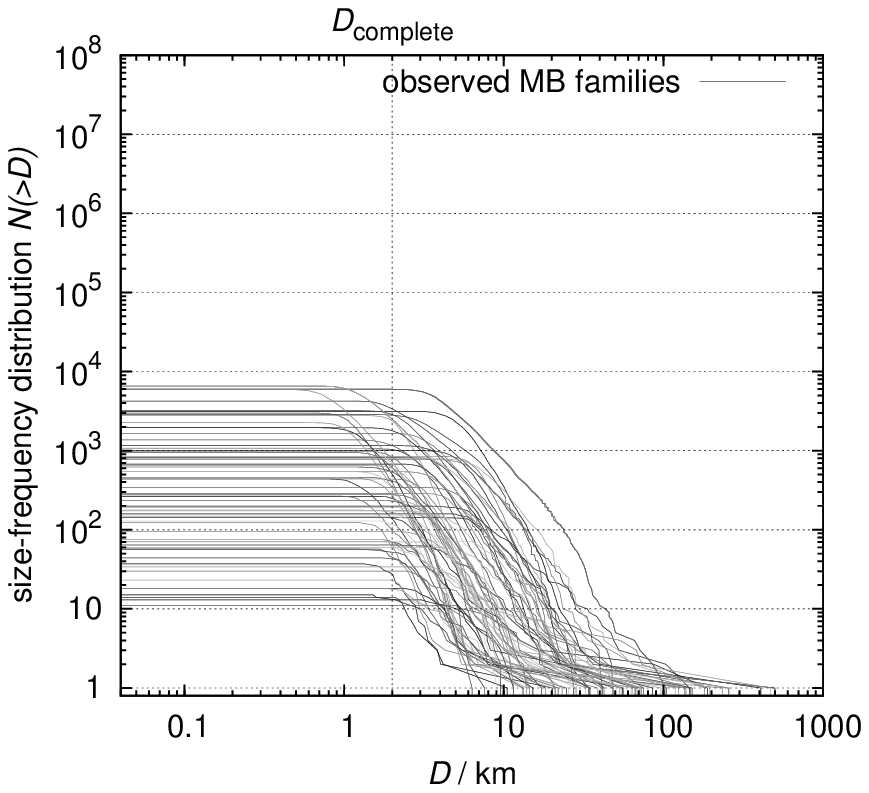}
\includegraphics[width=6cm]{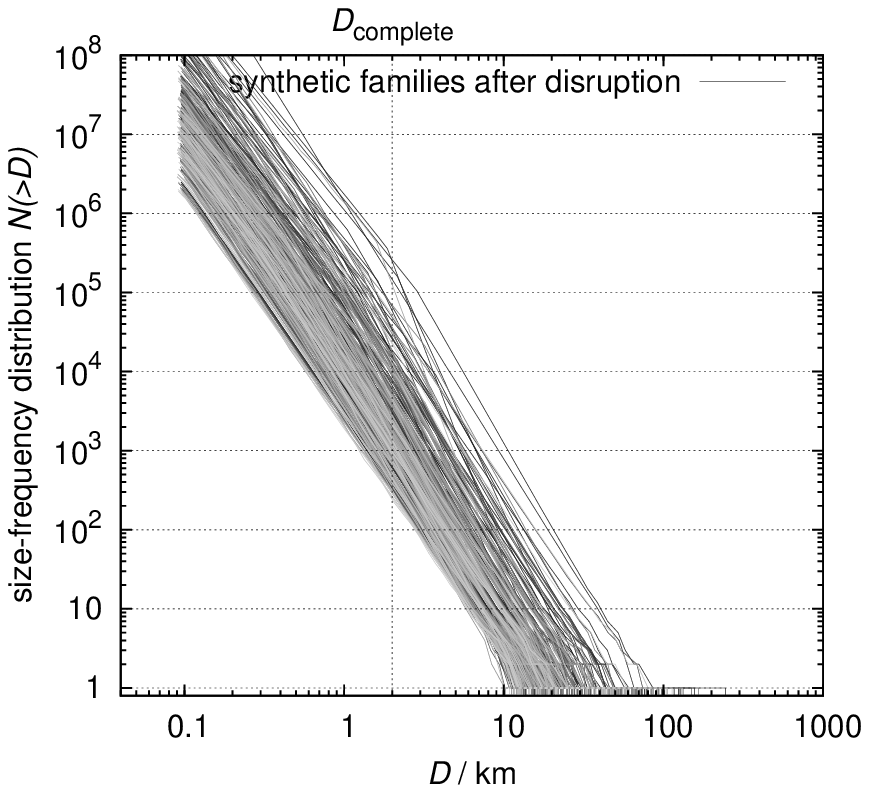}
\includegraphics[width=6cm]{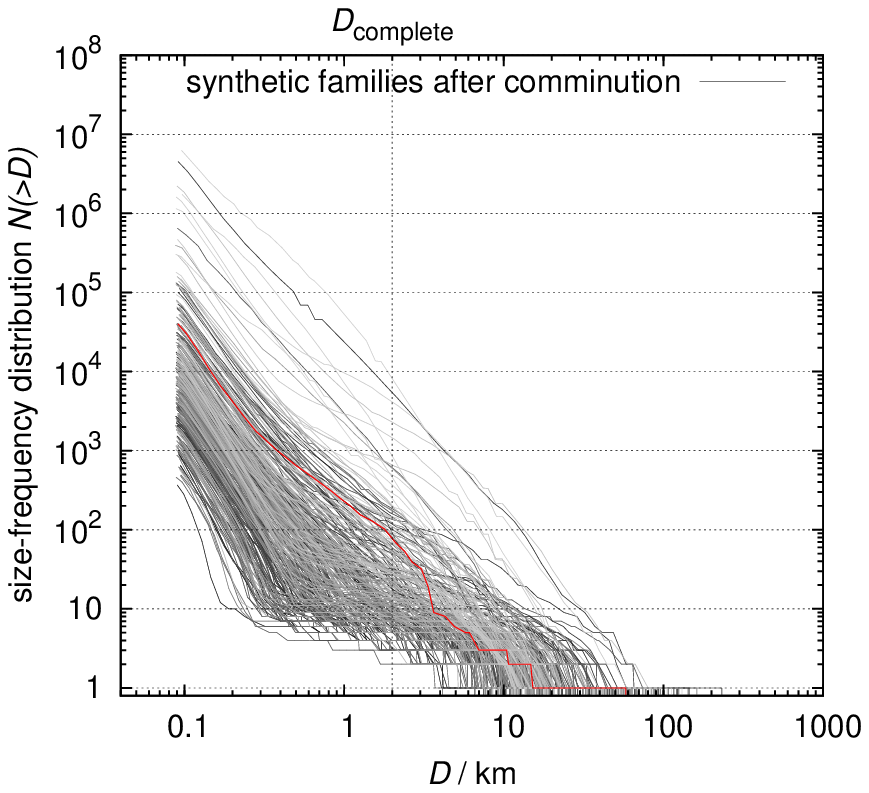}
\caption{Left panel: the size-frequency distributions of the observed asteroid families.
Middle panel: SFDs of 378 distinct synthetic families created during one of the collisional simulations of the MB and comets.
Initially, all synthetic SFDs are very steep, in agreement with SPH simulations (Durda et al. 2007).
We plot only the SFDs that fulfil the following criteria:
$D_{\rm PB} \ge 50$\,km,
$D_{\rm LF} \ge 10$\,km,
${\rm LR}/{\rm PB} < 0.5$ (i.e. catastrophic disruptions).
Right panel: the evolved SFDs after comminution.
Only a minority of families are observable now, since the number of remaining members
larger than the observational limit $D_{\rm limit} \simeq 2$\,km is often much smaller than~$100$.
The SFD that we use for the simulation in Section~\ref{sec:pristine} is denoted by red.}
\label{boulder_MB_comets_FAMS4_002_comminution_families_ob_SFD}
\end{figure*}

However, the situation in case of the LHB scenario is different.
Both the large population of comets and 
the several-times larger main belt, which has to withstand the cometary bombardment,
contribute to the enhanced comminution of the LHB families.
To estimate the amount of comminution, we performed the following calculations:
 i)~for a selected collisional simulation, whose production function is close to the average one,
    we recorded the SFDs of all synthetic families created in the course of time;
ii)~for each synthetic family, we restarted the simulation from the time~$t_0$
    when the family was crated until 4\,Gyr and saved the final SFD, i.e. after the comminution.
The results are shown in Figure~\ref{boulder_MB_comets_FAMS4_002_comminution_families_ob_SFD}.

It is now important to discuss criteria, which enable us to decide
if the comminutioned synthetic family would indeed be observable or not.
We use the following set of conditions:
$D_{\rm PB} \ge 50$\,km,
$D_{\rm LF} \ge 10$\,km (largest {\em fragment\/} is the first or the second largest body, where the SFD becomes steep),
${\rm LR}/{\rm PB} < 0.5$ (i.e. a catastrophic disruption).
Furthermore, we define $N_{\rm members}$ as the number of the {\em remaining\/} family members
larger than observational limit $D_{\rm limit} \simeq 2\,{\rm km}$
and use a condition $N_{\rm members} \ge 10$.
The latter number depends on the position of the family within the main belt, though.
In the favourable ``almost-empty'' zone (between $a_{\rm p} = 2.825\hbox{ and }2.955\,{\rm AU}$),
$N_{\rm members} \ge 10$ may be valid,
but in a populated part of the MB one would need $N_{\rm members} \gtrsim 100$ to detect the family.
The size distributions of synthetic families selected this way
resemble the observed SFDs of the main-belt families.

According to Figure~\ref{boulder_MB_REAL_SFD_FAMS4_sfd_4000} (3rd row),
where we can see the production functions after comminution for increasing values of~$N_{\rm members}$,
families with $D_{\rm PB} = 200$--400\,km remain {\em more prominent\/} than $D_{\rm PB} \simeq 100\,{\rm km}$ families
simply because they contain much more members with $D > 10\,{\rm km}$ that survive intact.
Our conclusion is thus that comminution may explain the paucity of the observed $D_{\rm PB} \simeq 100\,{\rm km}$ families.

\begin{figure*}
\centering
\includegraphics[width=7cm]{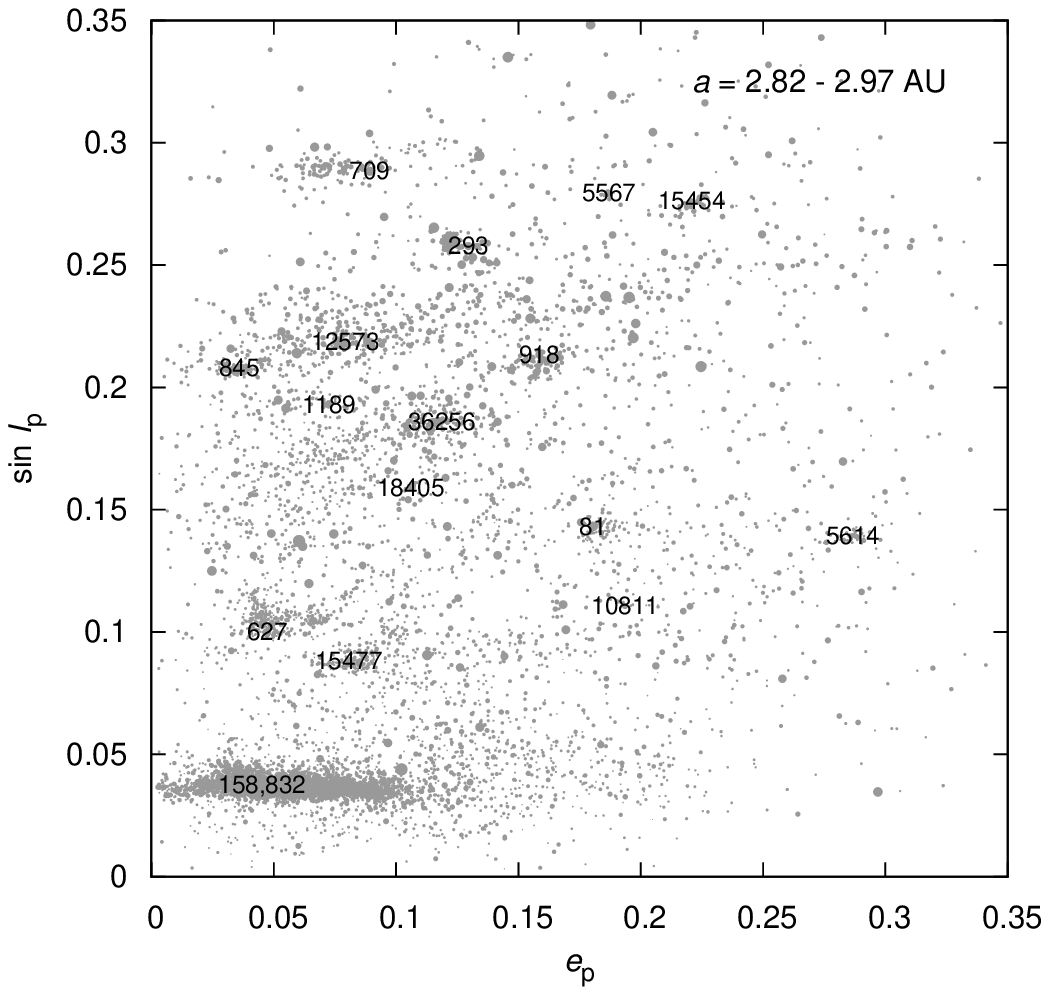}
\includegraphics[width=7cm]{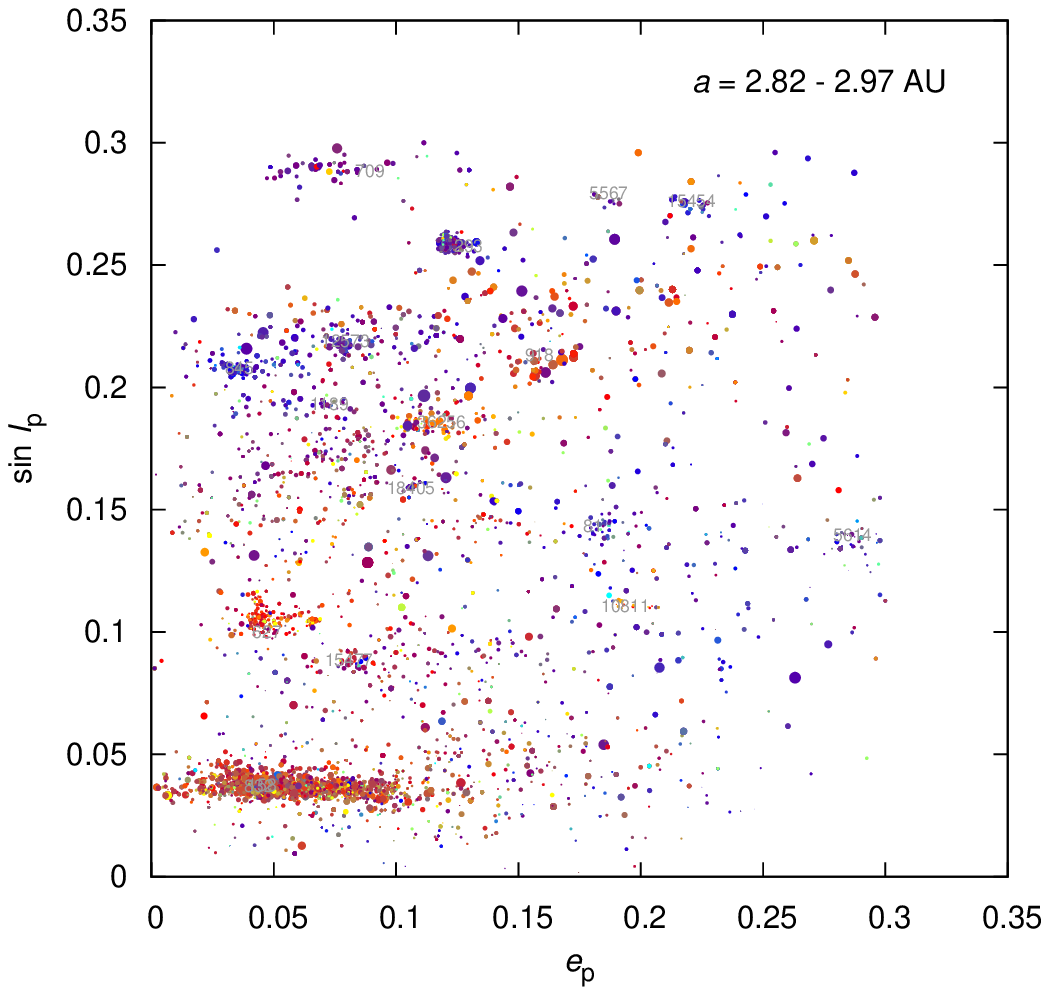}
\includegraphics[width=4cm]{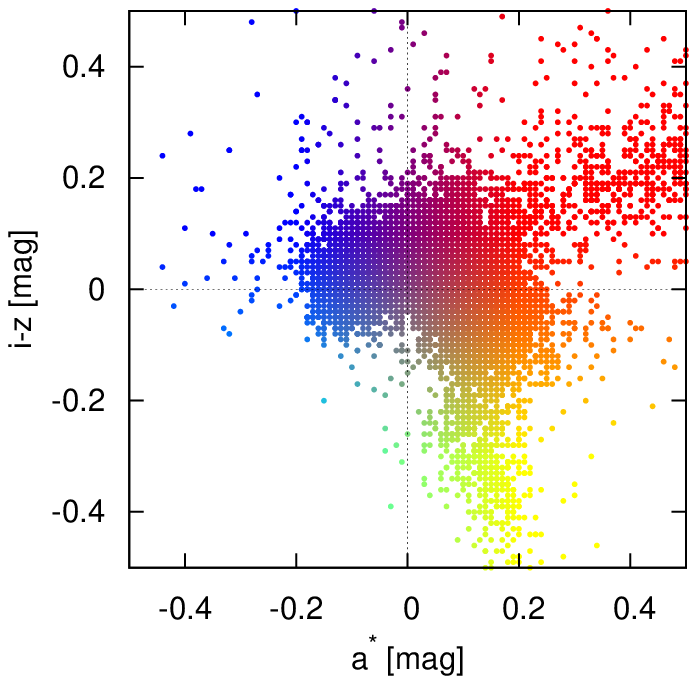}
\caption{The ``pristine zone'' of the main belt ($a_{\rm p} = 2.825\hbox{ to }2.955\,{\rm AU}$) displayed on the proper eccentricity~$e_{\rm p}$ vs proper inclination~$\sin I_{\rm p}$ plot.
Left panel: the sizes of symbols correspond to the sizes of asteroids, the families are denoted by designations.
Right panel: a subset of bodies for which SDSS data are available;
the colours of symbols correspond to the SDSS colour indices $a^*$ and $i-z$ (Parker et al. 2008).}
\label{pristine_MB_ei_SIZES_LABEL}
\end{figure*}

%%%%%%%%%%%%%%%%%%%%%%%%%%%%%%%%%%%%%%%%%%%%%%%%%%%%%%%%%%%%%%%%%%%%%%%%

\section{``Pristine zone'' between the 5:2 and 7:3 resonances}\label{sec:pristine}

We now focus on the zone between the 5:2 and 7:3 mean-motion resonances,
with $a_{\rm p} = 2.825\hbox{ to }2.955\,{\rm AU}$, which is not as populated
as the surrounding regions of the main belt (see Figure~\ref{families_ae}). This is a unique situation,
because both bounding resonances are strong enough to prevent any asteroids
from outside to enter this zone owing the Yarkovsky semimajor axis drift.
Any family formation event in the surroundings has only a minor influence
on this narrow region. It thus can be called ``pristine zone'' because it may
resemble the belt {\em prior\/} to creation of big asteroid families.

We identified nine previously unknown small families that are visible on the $(e_{\rm p}, \sin I_{\rm p})$~plot
(see Figure~\ref{pristine_MB_ei_SIZES_LABEL}). They are confirmed by the SDSS colours and WISE albedos, too.
Nevertheless, there is {\em only one\/} big and old family in this zone ($D_{\rm PB} \ge 100\,{\rm km}$),
i.e. Koronis.

%%%%%%%%%%%%%%%%%%%%%%%%%%%%%%%%%%%%%%%%%%%%%%%%%%%%%%%%%%%%%%%%%%%%%%%%

\begin{figure}
\centering
\includegraphics[width=6.5cm]{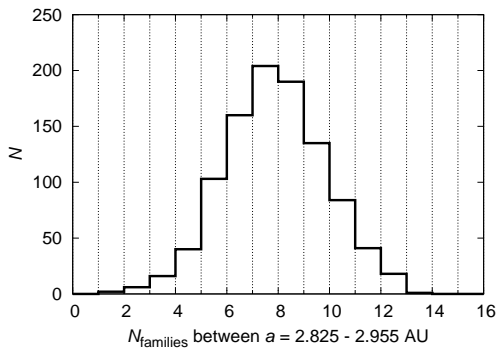}
\caption{The histogram for the expected number of LHB families located in the ``pristine zone'' of the main belt.}
\label{pristine_MB_probability}
\end{figure}

That at most one LHB family (Koronis) is observed in the ``pristine zone'' can give us
a simple probabilistic estimate for the {\em maximum\/} number of disruptions during the LHB.
We take the 192 existing main-belt bodies which have $D \ge 100\,{\rm km}$
and select randomly 100~of them that will break apart.
We repeat this selection 1000~times and always count the number
of families in the pristine zone. The resulting histogram is shown in Figure~\ref{pristine_MB_probability}.
As we can see, there is very low ($<$0.001) probability that the number
of families in the pristine zone is zero or one. On average we get eight families there,
i.e. about half of the 16 asteroids with $D \ge 100\,{\rm km}$ present in this zone.
It seems that either the number of disruptions should be substantially lower than 100
or we expect to find at least some ``remnants'' of the LHB families here.

%%%%%%%%%%%%%%%%%%%%%%%%%%%%%%%%%%%%%%%%%%%%%%%%%%%%%%%%%%%%%%%%%%%%%%%%

It is interesting that the SFD of an old comminutioned family is {\em very flat\/}
in the range $D = 1\hbox{ to }10\,{\rm km}$ (see Figure~\ref{boulder_MB_comets_FAMS4_002_comminution_families_ob_SFD})
--- similar to those of some of the ``less certain'' observed families!
We may speculate that the families like
%(709)~Fringilla,
(918)~Itha,
(5567)~Durisen,
(12573)~1999~NJ$_{53}$, or
(15454)~1998~YB$_3$ (all from the pristine zone)
are actually remnants of {\em larger and older\/} families, even though they are denoted as young.
It may be that the age estimate based on the $(a_{\rm p},H)$ analysis is incorrect
since small bodies were destroyed by comminution
and spread by the Yarkovsky effect too far away from the largest remnant,
so they can no longer be identified with the family.

%%%%%%%%%%%%%%%%%%%%%%%%%%%%%%%%%%%%%%%%%%%%%%%%%%%%%%%%%%%%%%%%%%%%%%%%

\begin{figure}
\centering
\includegraphics[width=7cm]{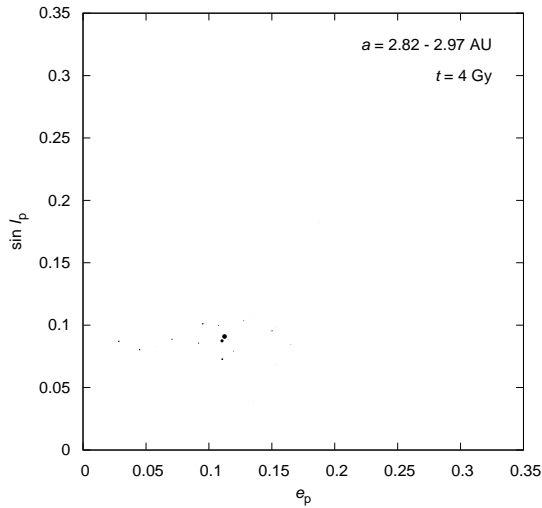}
\caption{The proper eccentricity vs proper inclination of
one synthetic old/comminutioned family evolved dynamically over 4\,Gyr.
Only a few family members ($N \simeq 10^1$) remained from
the original number of $N(D \!\ge\! 2\,{\rm km}) \simeq 10^2$.
The scales are the same as in Figure~\ref{pristine_MB_ei_SIZES_LABEL},
so we can compare it easily to the ``pristine zone''.}
\label{fams-2_COMMINUTION_ei_SIZES_COLOR_400My}
\end{figure}

Finally, we have to ask an important question:
what does an old/comminutioned family with $D_{\rm PB} \simeq 100\,{\rm km}$ look like in proper-element space?
To this aim, we created a synthetic family in the ``pristine zone'',
and assumed the family has $N_{\rm members} \simeq 100$ larger than $D_{\rm limit} \simeq 2\,{\rm km}$
and that the SFD is already flat in the $D = 1\hbox{ to }10\,{\rm km}$ range.
We evolved the asteroids up to 4\,Gyr due to the Yarkovsky effect
and gravitational perturbations, using the N-body integrator as in Section~\ref{sec:yarko_dispersal}.
Most of the $D \simeq 2\,{\rm km}$ bodies were lost in the course of the dynamical evolution, of course.
The resulting family is shown in Figure~\ref{fams-2_COMMINUTION_ei_SIZES_COLOR_400My}.
We can also imagine that this family is placed in the pristine zone among other observed families,
to get a feeling of whether it is easily observed or not (refer to Figure~\ref{pristine_MB_ei_SIZES_LABEL}).

It is clear that such family is {\em hardly observable\/} even in the almost empty zone of the main belt!
Our conclusion is that the comminution (as given by the Boulder code)
{\em can explain\/} the paucity of $D_{\rm PB} \simeq 100\,{\rm km}$ LHB families,
since we can hardly distinguish old families from the background.

%%%%%%%%%%%%%%%%%%%%%%%%%%%%%%%%%%%%%%%%%%%%%%%%%%%%%%%%%%%%%%%%%%%%%%%%

\section{Conclusions}\label{sec:conclusions}

In this paper we investigated the cometary bombardment of the asteroid
belt at the time of the LHB, in the framework of the Nice model. 
There is much evidence of a high cometary flux through the giant
planet region, but no strong evidence of a cometary bombardment on the
Moon. This suggests that many comets broke up on their way
to the inner solar system. By investigating the collisional evolution
of the asteroid belt and comparing the results to the collection
of actual collisional families, our aim was to constrain
whether the asteroid belt experienced an intense
cometary bombardment at the time of the LHB and, if possible,
constrain the intensity of this bombardment.  

Observations suggest that the number of collisional families is a very
shallow function of parent-body size (that we call in this paper
the ``production function''). We show that the collisional activity of
the asteroid belt as a closed system, i.e. without any external
cometary bombardment, in general does not produce such a shallow
production function. Moreover, the number of families with parent
bodies larger than 200\,km in diameter is in general too small compared
to the observations. However, there is a lot of stochasticity in the
collisional evolution of the asteroid belt, and about 5\,\% of our
simulations actually fit the observational constraints (shallowness of the
production function and number of large families) quite well. Thus,
in principle, there is no need for a bombardment due to external
agents (i.e. the comets) to explain the asteroid family collection,
provided that the real collisional evolution of the main belt was a
``lucky'' one and not the ``average'' one.

If one accounts for the bombardment provided by the comets crossing
the main belt at the LHB time, predicted by the Nice model, one can
easily justify the number of observed families with parent bodies larger
than 200\,km. However, the resulting production function is steep, and
the number of families produced by parent bodies of 100\,km is almost an
order of magnitude too large.

We have investigated several processes that may decimate the number of
families identifiable today with 100\,km parent bodies, without
considerably affecting the survival of families formed from larger
parent bodies. Of all these processes, the collisional comminution of
the families and their dispersal by the Yarkovsky effect are the most effective ones.
Provided that the physical disruption of comets due
to activity reduced the effective cometary flux through the
belt by a factor of $\approx 5$, the resulting distribution of families
(and consequently the Nice model) {\em is consistent\/} with observations.

\begin{figure}
\centering
\includegraphics[width=7cm]{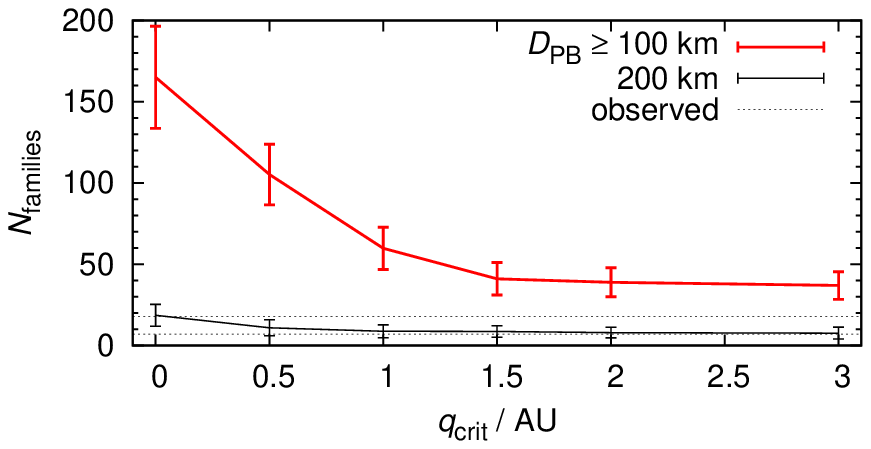}
\includegraphics[width=7cm]{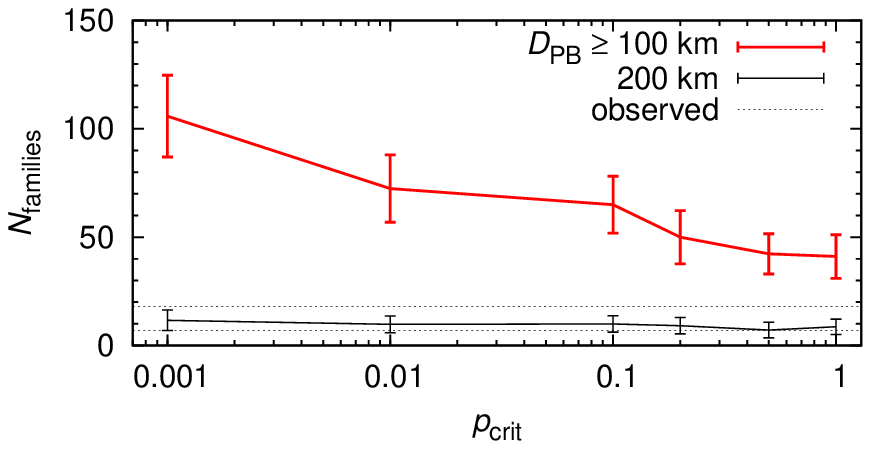}
\caption{The numbers of collisional families
for different critical perihelion distances~$q_{\rm crit}$ at which comets break up
and disruption probabilities~$p_{\rm crit}$ during one time step ($\Delta t = 500\,{\rm yr}$).
In the top panel, we vary~$q_{\rm crit}$ while keeping $p_{\rm crit} = 1$ constant.
In the bottom panel, $q_{\rm crit} = 1.5\,{\rm AU}$ is constant and we vary~$p_{\rm crit}$.
We always show the number of catastrophic disruptions
with parent-body sizes $D_{\rm PB} \ge 100\,{\rm km}$ (red line) and 200\,km (black line).
The error bars indicate typical (1-$\sigma$) spreads of Boulder simulations with different random seeds.
The observed numbers of corresponding families are indicated by thin dotted lines.}
\label{main_belt_qcrit}
\end{figure}

To better quantify the effects of various cometary-disruption laws,
we computed the numbers of asteroid families
for different critical perihelion distances~$q_{\rm crit}$
and for different disruption probabilities $p_{\rm crit}$ of comets during a given time step ($\Delta t = 500\,{\rm yr}$ in our case).
The results are summarised in Figure~\ref{main_belt_qcrit}.
Provided that comets are disrupted frequently enough,
namely the critical perihelion distance has to be at least $q_{\rm crit} \gtrsim 1\,{\rm AU}$,
while the probability of disruption is $p_{\rm crit} = 1$,
the number of $D_{\rm PB} \ge 100\,$km families drops by the aforementioned factor of $\approx 5$.
Alternatively, $q_{\rm crit}$ may be larger, but then comets have to
survive multiple perihelion passages (i.e. $p_{\rm crit}$ have to be lower than~1).
It would be very useful to test these conditions by {\em independent\/} models
of the evolution and physical disruptions of comets.
Such additional constraints on cometary-disruption laws would then enable
study of the original size-frequency distribution of the cometary disk in more detail.

We can also think of two ``alternative'' explanations:
 i)~physical lifetime of comets was strongly size-dependent
    so that smaller bodies break up easily compared to bigger ones;
ii)~high-velocity collisions between hard targets (asteroids)
    and {\em very\/} weak projectiles (comets) may result in different outcomes
    than in low-velocity regimes explored so far.
Our work thus may also serve as a motivation for further SPH simulations.

We finally emphasize that any collisional/dynamical models
of the main asteroid belt would benefit from the following advances:
\begin{itemize}
\item[i)] determination of reliable masses of asteroids of various classes.
This may be at least partly achieved by the Gaia mission in the near future.
Using up-to-date sizes and shape models (volumes) of asteroids one can then derive
their densities, which are directly related to ages of asteroid families.
\item[ii)] Development of methods for identifying asteroid families
and possibly targeted observations of larger asteroids addressing
their membership, which is sometimes critical for constructing
size-frequency distributions and for estimating parent-body sizes.
\item[iii)] An extension of the SHP simulations for both smaller and larger targets,
to assure that the scaling we use now is valid. Studies and laboratory measurements
of equations of states for different materials (e.g. cometary-like, porous)
are closely related to this issue.
\end{itemize}
The topics outlined above seem to be the most urgent developments
to be pursued in the future.

%%%%%%%%%%%%%%%%%%%%%%%%%%%%%%%%%%%%%%%%%%%%%%%%%%%%%%%%%%%%%%%%%%%%%%

\section*{Acknowledgements}

The work of MB and DV has been supported by the Grant Agency of the Czech
Republic (grants no.\ 205/08/0064, and 13-01308S)
and the Research Programme MSM0021620860 of the Czech Ministry of Education.
The work of WB and DN was supported by NASA's Lunar Science Institute
(Center for Lunar Origin and Evolution, grant number NNA09DB32A).
We also acknowledge the usage of computers of the Observatory and Planetarium in Hradec Kr\'alov\'e.
We thank Alberto Cellino for a careful review which helped to improve
final version of the paper.

%%%%%%%%%%%%%%%%%%%%%%%%%%%%%%%%%%%%%%%%%%%%%%%%%%%%%%%%%%%%%%%%%%%%%%

%%%%%%%%%%%%%%%%%%%%%%%%%%%%%%%%%%%%%%%%%%%%%%%%%%%%%%%%%%%%%%%%%%%%%%%%
%%%%%%%%%%%%%%%%%%%%%%%%%%%%%%%%%%%%%%%%%%%%%%%%%%%%%%%%%%%%%%%%%%%%%%%%

%\appendix

%%%%%%%%%%%%%%%%%%%%%%%%%%%%%%%%%%%%%%%%%%%%%%%%%%%%%%%%%%%%%%%%%%%%%%%%
%%%%%%%%%%%%%%%%%%%%%%%%%%%%%%%%%%%%%%%%%%%%%%%%%%%%%%%%%%%%%%%%%%%%%%%%

\label{lastpage}

\end{document}